\documentclass[sn-nature]{sn-jnl}

\usepackage{comment}
\usepackage[normalem]{ulem}

\def\bzeta{{\mbox{\boldmath $\zeta $}}}

\usepackage[normalem]{ulem}

\usepackage{float}
\usepackage{arydshln}

\usepackage{pdfpages}

\usepackage{graphicx}%
\usepackage{multirow}%
\usepackage{amsmath,amssymb,amsfonts}%
\usepackage{amsthm}%
\usepackage{mathrsfs}%
\usepackage[title]{appendix}%
\usepackage{xcolor}%
\usepackage{textcomp}%
\usepackage{manyfoot}%
\usepackage{booktabs}%
\usepackage{verbatim}
\usepackage{placeins}
\usepackage{hyperref} 

\usepackage{bm}

\usepackage[left]{lineno}

\hypersetup{
    colorlinks=true,
    linkcolor=blue,
    bookmarksopen=true,
    bookmarksopenlevel=3
}

\def\bxi{{\mbox{\boldmath $\xi $}}}

\renewcommand{\vec}[1]{\mathbf{#1}}
\newcommand{\uvec}[1]{\mathbf{\hat{#1}}}

\usepackage[normalem]{ulem}
\def\bzeta{{\mbox{\boldmath $\zeta $}}}
\usepackage[normalem]{ulem}

\usepackage{float}

\usepackage{caption}  
\begin{document}

\title[Article Title]{Dissipative particle dynamics models of encapsulated microbubbles and gas vesicles for biomedical ultrasound simulations}

\author[1,2]{\fnm{Nikolaos} \sur{Ntarakas}}\email{nikolaos.ntarakas@ki.si}
\author[1,2]{\fnm{Maša} \sur{Lah}}\email{masa.lah@ki.si}
\author[1,2]{\fnm{Daniel} \sur{Sven\v{s}ek}}\email{daniel.svensek@fmf.uni-lj.si}
\author[1,2]{\fnm{Tilen} \sur{Potisk}}\email{tilen.potisk@ki.si}
\author*[1,2,3]{\fnm{Matej} \sur{Praprotnik}}\email{praprot@cmm.ki.si}

\affil[1]{\orgdiv{Laboratory for Molecular Modeling}, \orgname{National Institute of Chemistry}, \orgaddress{\street{Hajdrihova 19}, \postcode{SI-1001} \city{Ljubljana}, \country{Slovenia}}}

\affil[2]{\orgdiv{Department of Physics}, \orgname{Faculty of Mathematics and Physics, University of Ljubljana}, \orgaddress{\street{Jadranska 19}, \postcode{SI-1000} \city{Ljubljana}, \country{Slovenia}}}

\affil[3]{\orgdiv{Universitat de Barcelona Institute of Complex Systems (UBICS)},
\newline
\orgaddress{\street{C/ Martí i Franqués 1}, \postcode{08028} \city{Barcelona}, \country{Spain}}}

\abstract{Ultrasound-guided drug and gene delivery (\textsc{usdg}) enables controlled and spatially precise delivery of drugs and macromolecules, encapsulated in microbubbles (\textsc{emb}s) and submicron gas vesicles (\textsc{gv}s), to target areas such as cancer tumors. It is a non-invasive, high precision, low toxicity process with drastically reduced drug dosage. Rheological and acoustic properties of \textsc{gv}s and \textsc{emb}s critically affect the outcome of \textsc{usdg} and imaging. Detailed understanding and modeling of their physical properties is thus essential for ultrasound-mediated therapeutic applications. 
State-of-the-art continuuum models of shelled bodies
cannot incorporate critical details such as varying thickness of the encapsulating shell or specific interactions between its constituents and interior or exterior solvents. 
Such modeling approaches also do not allow for detailed modeling of chemical surface functionalizations, which are crucial for tuning the \textsc{gv}--blood interactions.
We develop a general particle-based modeling framework for encapsulated bodies that accurately captures elastic and rheological properties of \textsc{gv}s and \textsc{emb}s. We use dissipative particle dynamics to model the solvent, the gaseous phase in the capsid, and the triangulated surfaces of immersed objects. Their elastic behavior is studied and validated through stretching and buckling simulations, eigenmode analysis, shear flow simulations, and comparison of predicted \textsc{gv} buckling pressure with experimental data from the literature.
The presented modeling approach paves the way for large-scale simulations of encapsulated bodies, capturing their dynamics, interactions, and collective behavior.}


\maketitle

\section*{Introduction}
\addcontentsline{toc}{section}{Introduction}

\noindent
Ultrasound (\textsc{us}) is increasingly being used in biomedical applications to diagnose many types of cancer, for blood flow analysis and therapeutic applications, including thermal tissue coagulation, kidney stones fragmentation, bone healing, mechanical tissue disruption and in cases of joint inflammation or rheumatoid arthritis \cite{lee2017,lindner2004,schroeder2009b,mace2011,tanter2014,rungta2017,Mo:2012coussios}. It offers numerous advantages, such as functionality in opaque media, relatively high spatial precision on the micrometer scale and fast, reconfigurable field formation \cite{wu2023}. These features have made \textsc{us} a cornerstone of modern biomedical imaging and therapy.

To further enhance the capabilities of \textsc{us} in diagnostics and therapeutics, encapsulated biomaterials \cite{lindner2004,Paul:2014} have emerged as powerful tools. Their unique design and shell properties make them highly adaptable for two major \textsc{us} biomedical applications: enhancing \textsc{us} imaging as ultrasound contrast agents (\textsc{uca}s) \cite{lee2017,tanter2014,Paul:2014,wang2014a} and enabling the encapsulation and targeted delivery of therapeutic drugs \cite{lee2017,lindner2004,Paul:2014,wang2014a,Unger:2014,Meng:2019,Konofagou:2009,Konofagou:2011,Konofagou:2012}. Among these materials, encapsulated microbubbles (\textsc{emb}s) and gas vesicles (\textsc{gv}s) have garnered significant attention for their adaptability and effectiveness in such applications. Radial oscillations of \textsc{emb}s and \textsc{gv}s generate strong nonlinear acoustic signals with a unique signature in the acoustic field and a frequency range much greater than that produced by tissues. This allows them to generate significant \textsc{us} contrast across a range of frequencies, supporting harmonic, multiplexed, and multimodal \textsc{us} imaging, as well as cell-specific molecular targeting \cite{lakshmanan2016,maresca2017}.

\textsc{emb}s injected into the bloodstream are already being used for echocardiography \cite{Main:2014,Paul:2014,Unger:2014}, which is one of the essential tools for diagnosing cardiovascular diseases. \textsc{emb}s are typically $1$\,\textmu m in diameter and are comprised of biologically inert gases, such as air or a gas with lower water solubility, stabilized within a lipid, protein, or polymer shell \cite{wang2014a,lee2017}. 
When an \textsc{emb} is subjected to a high-intensity acoustic field, it expands in volume and collapses violently. This process is known as inertial cavitation \cite{Choi:2012coussios}. In contrast, during non-inertial or stable cavitation, \textsc{emb}s oscillate with relatively minor deformations under lower acoustic pressure amplitudes. Cavitation is utilized in sonoporation \cite{Coussios:2008}, which is a targeted drug delivery technique that creates temporary pores in cell membranes, enabling the entry of foreign substances \cite{Pandur:2023,Meng:2019}. Exploiting the sonoporation effect for disease therapy has many advantages, for instance, injecting \textsc{emb}s intravenously can lower drug dosage and minimize side effects
of nonspecific drug delivery into healthy organs as the \textsc{emb}s only collapse in specific diseased areas due to focused \textsc{us} irradiation.
One limitation to the use of \textsc{emb}s is their size that prevents them from extravasating into tumors. To circumvent this limitation, \textsc{gv}s have been introduced as a new class of nanoscale \textsc{us} imaging agents  \cite{ezzeldin2012,park2017a,walsby1972a,walsby1994,walsbyanthonyedward1971a,xu2014,yang2017}.

\textsc{gv}s are gas-filled, protein-shelled nanostructures derived from buoyant photosynthetic microbes. These vesicles vary in size, with widths ranging from $45$ to $200$\,nm and lengths from $100$ to $800$\,nm, depending on their genetic origin, and can undergo external pressures of several bar without collapsing. Recently, \textsc{gv}s have also been shown to increase the influx of calcium ions, when attached to biological cells and insonated by ultrasound \cite{bencina2024}. The structure of several types of \textsc{gv}s has already been characterized using cryo-EM (Cryogenic electron microscopy) and cryo-ET (Cryogenic electron tomography) \cite{shapiro2023, huber2023}, revealing that the main structural protein GvpA self-assembles helically in a cylindrical shape, which closes off on both sides by cone-shaped tips. The polarity of the helical assembly inverts at the midpoint of the \textsc{gv} cylinder, which may act as an elongation center for growth. This implies that the ribs are oriented helically along the cylinder, reversing direction at the central rib \cite{Offner:1998,shapiro2023}. Unlike \textsc{emb}s, which confine preloaded gas in an unstable state, \textsc{gv}s have $2$ nm-thick protein shells that exclude water but allow gas to diffuse freely in and out of their interior \cite{maresca2018}. 

The acoustic behavior of \textsc{emb}s and \textsc{gv}s is influenced by several factors, such as the viscosity and temperature of the surrounding fluid, the applied acoustic pressure and the physical characteristics of the \textsc{emb}s, including size and shell properties like viscosity and elasticity \cite{jong2002,sboros2008}. In addition, the presence of nearby vessel walls or cells can significantly affect the \textsc{emb} behavior \cite{sboros2008}.
Unlike the detailed modeling of blood flow \cite{noguchi2005a,dupin2007,discher1998,li2005,pivkin2008,gompper1997,gompper2004,noguchi2005b,fedosov2010,fedosov2010a,fedosov2011,fedosov2014a,muller2016,shi2013,ye2014a,ye2014} or cloud cavitation collapse \cite{rossinelli2013,Rasthofer:2019}, the current theoretical modeling of \textsc{emb} oscillations primarily relies on the continuum theory developed by Rayleigh and Plesset for a single, free, spherically symmetric bubble in an infinite liquid with constant viscosity \cite{f.r.s1917,hickling1964}. 
The Rayleigh-Plesset \textsc{emb} model incorporates several assumptions, including the ideal gas behavior of the encapsulated gas and the absence of a shell. A series of increasingly complex models have been developed to more accurately represent the dynamics of \textsc{emb}s in vivo — 
 particularly those excited by \textsc{us} while flowing through small blood vessels. Despite these improvements, the models continue to rely on various assumptions and simplifications \cite{dollet2019,medwin1977a,qin2009a}.

Continuum models for \textsc{gv}s are very scarce, with the exception of finite element models for various types of \textsc{gv}s, such as the \textit{Anabaena flos-aquae} \cite{maresca2017,zhang2020vibration,shapiro2022,shapiro2023} and the \textit{Halobacterium salinarum} \cite{cherin2017}, which focus purely on mechanical properties in vacuum, without explicitly modeling the surrounding solvent or encapsulated gas. A microscopic model has been reported \cite{patanakar2023}, in which a model for the GvpA rib was developed and used to calculate the Young's moduli of the \textsc{gv} shell.
Although significant efforts have been made to improve continuum models for \textsc{emb}s and \textsc{gv}s \cite{dollet2019,dejong1992,dejong1994b,dejong1994c,jong2002,marmottant2005,marmottantphilippe2008,marmottant2011,sassaroli2005,hosseinkhah2012,hosseinkhah2015,qin2009a,cherin2017}, accurately modeling the shell properties before and after insonation remains a challenging task. The applicability of continuum models in these scenarios is limited, mainly due to the lack of detailed interfacial constitutive models \cite{dollet2019}. These limitations of existing models based on continuum theory preclude an accurate description of cavitation, drastically degrading the prediction of drug delivery outcomes.

The development of novel \textsc{emb}s and \textsc{gv}s models using mesoscopic particle-based approaches tailored to the specific shell material is crucial to study changes in the material upon deformation and its mechanical response to interaction with \textsc{us}. Importantly, the mechanical behavior of \textsc{uca}s differs significantly between water and blood, due to variations in viscosity, elasticity, and the complex interplay with surrounding \textsc{emb}s and vesicles in the bloodstream. Incorporating these factors into simulations is essential for accurate predictions of their performance in real physiological environments.
To accurately capture the rheological and acoustic properties, as well as the dynamics of \textsc{emb}s and \textsc{gv}s, we propose mesoscopic particle-based models inspired by the network models of red blood cells (\textsc{rbc}s). The proposed models are designed to be general enough to accommodate a wide range of physical systems.
Here, we use the dissipative particle dynamics (\textsc{dpd}) method, a state-of-the-art particle-based method for modeling colloidal suspensions, polymers, soft matter, and simple fluids. We model mechanical properties of \textsc{emb}s and \textsc{gv}s, including their behavior in stretching and buckling experiments. Our predictions for the buckling pressure of \textsc{gv}s are compared to experimental measurements. Furthermore, we determine fundamental eigenmodes of \textsc{emb} and \textsc{gv}s. Finally, we study their rheological properties under shear flow and compare them with analytical expressions.

\section*{Results}
\addcontentsline{toc}{section}{Results}
\label{sec: Results}
\noindent

\subsection*{General particle-based elasticity framework for simulating membrane-encapsulated soft- and biomaterials}
\addcontentsline{toc}{subsection}{General particle-based elasticity framework for simulating membrane-encapsulated soft- and biomaterials}

\noindent
The role of membranes in soft- and biomaterials is multifaceted. For biological cells, the membrane separates the interior from external disturbances and can also provide means for the exchange of ions, solvents, gas molecules, and other substances \cite{Ussing:1949}. Biological membranes typically consist of lipid, polymer, or protein units. The type of these units, their interaction with each other and the environment, as well as their binding topology dictate the elastic behavior of membranes. 

\begin{figure}
\centering
    \includegraphics[width=\textwidth]{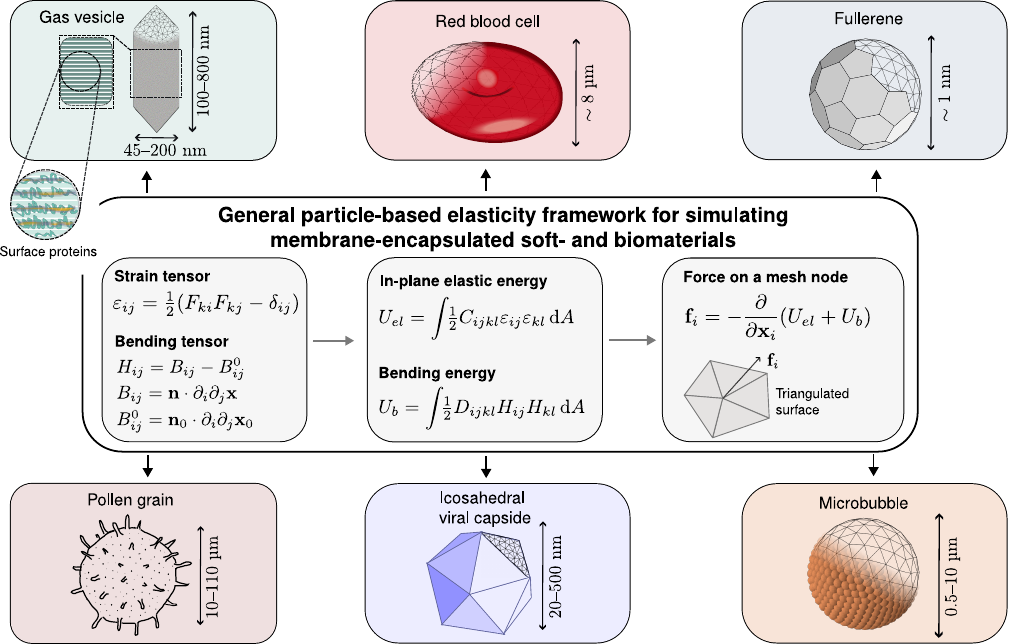}
\caption{Overview of particle-based modeling of elastic behavior of thin shells used in this work. 
An object is described by its subunits and a triangulated network. Elastic energy is split into two parts: in-plane elastic energy, characterized by the material elastic tensor $C_{ijkl}$ and the strain tensor $\varepsilon_{ij}$ measuring the deformations; and bending energy, characterized by the flexural rigidity tensor $D_{ijkl}$ and the bending tensor $H_{ij}$. 
The corresponding vertex forces are calculated by taking 
derivatives of the total elastic energy with respect to vertex positions. 
The modeling framework is applicable to biomaterials of  different shapes and arbitrary local anisotropy.}
\label{fig: general}
\end{figure}

This extreme diversity of membranes with different elastic properties requires a general methodology that is capable of incorporating various possible symmetries, for example, the isotropic elasticity of \textsc{emb}s on one hand, and orthotropic elasticity of \textsc{gv}s on the other. There are different approaches to modeling elastic properties of thin shells. A popular one is connecting the various subunits with harmonic bonds or potential wells, aided by a harmonic angular potential \cite{pandit2008simulations}. It is well known that a straightforward application of such an approach --- using harmonic bonds with equal spring constants --- does not lead to general elastic behavior. An alternative approach, which we follow here, is to discretize the continuum elastic surface energy on a triangulated surface spanned by the various subunits (see Fig.~\ref{fig: general}) \cite{Lidmar:2003}. The continuum energy expression is then used in the subsequent per-vertex force calculations. 

We quantify the deformation using the deformation gradient $F_{ij} = \frac{\partial x_i}{\partial X_j}$, which relates the difference in positions of two infinitesimally close material points in the deformed configuration $\mathbf{x}$ to their difference in the reference configuration $\mathbf{X}$ \cite{Bonet:1997}. 
To exclude local rotation, deformation is typically described using the Green-Lagrange deformation tensor (strain tensor) $\bm{\varepsilon} = \frac12(\bm{\mathsf{F}}^T\bm{\mathsf{F}}-\bm{\mathsf{I}})$, which measures the deformation relative to the reference configuration. Moreover, we incorporate anisotropic elasticity 
to model
the diverse and generally complex elastic properties of biological membranes.
Anisotropic objects are characterized by their reduced symmetry group or equivalently by a set of structural tensors $\bm{\mathsf{M}}$, which reflect the distinguished directions, lines, or planes of an object. See Supplementary Information~\ref{appendix:b} for a detailed overview of the effects of material symmetry and the principle of isotropy of space on the form of the elastic energy.

In linear elastic theory of thin shells, in-plane and bending deformations are decoupled, and the total elastic energy is $U=U_{el} + U_b$, where $U_{el}$ and $U_b$ are the in-plane elastic and bending energies, respectively.

The general in-plane elastic energy is formulated using the two-dimensional (2D) in-plane strain tensor $\bm{\varepsilon}$,
\begin{equation} \label{Uel-general}
    U_{el}=\int\!\! {\textstyle\frac12} C_{ijkl}\varepsilon_{ij}\varepsilon_{kl}\,\mathrm{d}A ,
\end{equation}
where ${\rm d}A$ is the surface element of the shell, and $C_{ijkl} = h\, C_{ijkl}^{3D}$ is its in-plane elastic tensor, derived from the three-dimensional (3D) material elastic tensor $C_{ijkl}^{3D}$ and the shell thickness $h$, Eq.\,(\ref{cijkla}) in Supplementary Information \ref{appendix:b}.
The in-plane stress tensor $\sigma_{ij} = C_{ijkl}\varepsilon_{kl}$ has units of force per unit length.
The tensor $C_{ijkl}$ satisfies $C_{ijkl}=C_{jikl}=C_{ijlk}$, which reflects the symmetries of the strain and stress tensors. Under the assumption of hyperelasticity, where stress is derived from an elastic potential, it also satisfies $C_{ijkl}=C_{klij}$.
These symmetries reduce the maximum possible number of elastic parameters from $81$ to $21$ in 3D and from $16$ to $6$ in 2D.
The in-plane elastic tensor $C_{ijkl}$ is expressed as a combination of Kronecker delta $\delta_{ij}$ and in-plane structural tensor(s) $M_{ij}$, as derived in Supplementary Information \ref{appendix:b} in the context of orthotropic elasticity of \textsc{gv}s.

A general form of the bending energy is
\begin{equation} \label{bend}
    U_{b} = \int\!\!{\textstyle\frac12} D_{ijkl}H_{ij}H_{kl}\,\mathrm{d}A,
\end{equation}
where $H_{ij} = B_{ij} - B_{ij}^0$ 
is the bending tensor \cite{Komura2005}, which measures the deviation of the curvature tensor 
$B_{ij} = \mathbf{n}\cdot\partial_i \partial_j \mathbf{x}$ 
in the deformed state from the spontaneous curvature tensor 
$B_{ij}^0 = \mathbf{n}_0\cdot\partial_i \partial_j \mathbf{x}_0$.
Here, 
$\mathbf{n}$ and $\mathbf{n}_0$
denote the normals to the deformed ($\bf x$) and undeformed (${\bf x}_0$) configuration surfaces, respectively. 
The principal directions of the curvature tensor align with extremal curvatures, represented by its eigenvalues $1/R_1$ and $1/R_2$, where $R_1$ and $R_2$ are the radii of curvature.
In linear thin shell elasticity, the material flexural rigidity tensor $D_{ijkl}$ is fully specified by the material elastic tensor through the relation $D_{ijkl} = \frac{h^3}{12}\,C_{ijkl}^{3D} = \frac{h^2}{12}\,C_{ijkl}$, Eq.\,(\ref{dijkla}) in \ref{appendix:b}. 

In the constant strain triangle approximation (CST) \cite{Turner:1956}, where the strain field $\bm{\varepsilon}$ is assumed to be constant within each triangle, the surface integrals Eqs.\,(\ref{Uel-general})-(\ref{bend}) can be replaced by sums over the triangles of the triangulated surface, as given in Eqs.\,(\ref{inplane}) and (\ref{Ub-t}) of Supplementary Information.

\subsection*{Models of ultrasound contrast agents}
\addcontentsline{toc}{subsection}{Models of ultrasound contrast agents}
\noindent
Within the introduced general elastic particle-based computational framework, we focus on modeling the behavior of two encapsulated agents:  \textsc{emb}s and \textsc{gv}s.
The specific models we employ are inspired by \textsc{rbc} membrane models \cite{noguchi2005a,noguchi2005b,li2005,dupin2007,pivkin2008,Fedosov:2010,fedosov2010a,fedosov2011,fedosov2014a,ye2014, ye2014a,economides2017,economides2021,lucas2023}. The \textsc{rbc} membrane consists of two main components: a pseudo-hexagonal elastic spectrin network \cite{liu1987}, and a fluid-like lipid bilayer. In contrast, polymer- or protein-based \textsc{emb}s and \textsc{gv}s comprise only an elastic network \cite{walsby1994}, while lipid-based microbubbles are encapsulated by a lipid monolayer membrane \cite{marmottant2005}. 
The primary differences in modeling \textsc{emb}s and \textsc{gv}s compared to \textsc{rbc}s lie in the different topology of the triangulated surfaces and, in the case of \textsc{gv}s, the inclusion of anisotropic elastic terms.

\subsubsection*{Microbubbles}
\addcontentsline{toc}{subsubsection}{Microbubbles}
\label{sec: MBs}

\noindent
\textsc{emb} shells are made of proteins, polymers, or lipids. Most \textsc{emb}s appear to be well described by an isotropic elastic model \cite{Vlachomitrou:2017}, although there are continuum models that assume transversely isotropic elastic shells where the anisotropy axis is along the radial direction \cite{chabouh2023}.

The elastic tensor $C_{ijkl}$ of isotropic materials is exclusively expressed through the isotropic tensor---the Kronecker delta $\delta_{ij}$. It has two independent terms:
\begin{equation} \label{isotensor}
    C_{ijkl} = K_a \delta_{ij}\delta_{kl} + \mu\left(\delta_{ik}\delta_{jl}+\delta_{il}\delta_{jk}-\delta_{ij}\delta_{kl}\right),
\end{equation}
where $K_a$ and $\mu$ are the bulk and the shear moduli, respectively. There is a more detailed description of the elastic moduli and the expression for elastic energy in Supplementary Information \ref{appendix:b}.

The bending energy follows from Eq.\,(\ref{bend}) and, for isotropic shells, consists of two independent terms:
\begin{equation} \label{iso-bend}
    U_b = \int\left[{\textstyle\frac{1}{2}}\kappa J_1^2 - \kappa (1-\nu) J_2\right]\mathrm{d}A,
\end{equation}
where $\kappa = \frac{Eh^2}{12(1-\nu^2)}$ is the bending constant, $E$, $\nu$ are the 2D Young's modulus and Poisson's ratio (Eqs.\,(\ref{Eeng})-(\ref{nueng}) in \ref{appendix:b}), and the scalar differential curvature invariants are defined as 
$J_1 = \mathrm{tr}(\bm{\mathsf{H}})$ and $J_2 = \frac12[\mathrm{tr}(\bm{\mathsf{H}})^2 - \mathrm{tr}(\bm{\mathsf{H}}^2)] = \mathrm{det}(\bm{\mathsf{H}})$. 
The bending energy Eq.\,(\ref{iso-bend}) is discretized using the Kantor-Nelson approach, see Supplementary Information \ref{appendix:b}.

\subsubsection*{Gas vesicles}
\addcontentsline{toc}{subsubsection}{Gas vesicles}
\label{gv_section}
\noindent
The \textsc{gv} membrane consists of GvpA protein ribs arranged helically around the \textsc{gv} axis \cite{Offner:1998,shapiro2023}, see Fig.~\ref{fig: general}. Its elastic properties differ along the ribs and perpendicular to them \cite{shapiro2022,shapiro2023}. 
The local material symmetry group $\mathcal{G}$ is therefore orthotropic, spanned by four elements: identity, inversion and a pair of reflections across the rib direction and perpendicular to it.
The 2D elastic tensor can therefore be constructed using Kronecker delta $\delta_{ij}$ and one structural tensor, which is invariant to all group elements in $\mathcal{G}$: $\bm{\mathsf{M}} = \mathbf{m}\otimes \mathbf{m}$, where $\mathbf{m}$ is chosen to point perpendicular to the rib. Since the ribs run nearly perpendicular to the \textsc{gv} axis, $\bf m$ is well approximated by a projection (see Supplementary Information \ref{secA1}).

Taking into account this symmetry, the most general form of the elastic tensor is
\begin{align} \label{tensor}
    C_{ijkl} &= K_a\, \delta_{ij}\delta_{kl} + \mu\left(\delta_{ik}\delta_{jl}+\delta_{il}\delta_{jk}-\delta_{ij}\delta_{kl}\right) \nonumber\\ 
    &+ (\mu_L-\mu)\,(m_im_l \delta_{jk} + m_j m_l \delta_{ik} + m_i m_k \delta_{jl} + m_jm_k \delta_{il}) \nonumber\\
    &+ c\, m_im_jm_km_l,
\end{align}
with $\mu_L>0$, $c>0$ \cite{Lempriere:1968,Li:2014ortho} 
the anisotropic elastic coefficients, which are positive for stability reasons. 
The coefficient $\mu_L$ is the membrane's (in-plane) shear elastic constant, while $c$ contributes to the stiffness along the anisotropy axis $\mathbf{m}$ \cite{spencer1984}. A more in-depth explanation regarding the in-plane elastic energy is provided in Supplementary Information \ref{appendix:b}. 
In the linear regime, the coefficients $K_a, \mu, \mu_L$, and $c$ in Eq.\,(\ref{tensor}) can be related to the engineering constants: Young's modulus along the \textsc{gv} axis, $E_l$, and perpendicular to it (along the ribs), $E_t$, Poisson's ratio $\nu_{lt}$ for stretching along the ribs, and shear modulus $G (= \mu_L)$. The other Poisson's ratio $\nu_{tl}$ is already fixed with these choices.
These relations are given in Eqs.\,(\ref{Kmu-eng})-(\ref{eng-Kmu}) of \ref{appendix:b}.

Following Eq.\,(\ref{bend}), the bending energy of a thin orthotropic shell can be expressed as a sum of four independent terms:
\begin{equation} \label{gv-bend-ene}
    U_b = \int\left({\textstyle\frac{1}{2}}\kappa_tJ_1^2 - \kappa_t(1-\nu_{lt}) J_2 + \kappa_\mu J_3 + {\textstyle\frac12} \kappa_c J_4^2\right)\mathrm{d}A,
\end{equation}
where new scalar differential curvature invariants are defined as $J_3 = \mathbf{m}^T\bm{\mathsf{H}}^T\bm{\mathsf{H}} \mathbf{m}$, $J_4 = \mathbf{m}^T\bm{\mathsf{H}} \mathbf{m}$, and 
$\kappa_t = \frac{E_th^2}{12(1-\nu_{lt}\nu_{tl})}$, $\kappa_\mu = \frac{(\mu-\mu_L)h^2}{6}$, $\kappa_c = \frac{ch^2}{12}$ 
are the bending constants in the thin shell regime \cite{Timoshenko:1959}.

For simplicity, we keep the bending energy of the \textsc{gv} membrane isotropic in this work and set the bending constant to $\kappa = \frac{E_th^2}{12(1-\nu_{lt}^2)}$, where we have used the smaller Young's modulus $E_t$. This is to ensure that the circumferential instability, which in buckling experiments occurs at lower pressure amplitudes than other more complicated instabilities, has the correct energy cost.

\subsection*{Mechanical properties}
\addcontentsline{toc}{subsection}{Mechanical properties}
\noindent
Mechanical properties of biomaterials can be determined by performing elementary mechanical experiments, such as stretching \cite{Dao:2003}, torsion \cite{Rivlin:1951}, and compression \cite{Stammen:2001}. 
Stretching experiments are typically done using optical tweezers, where two micrometer-sized spherical silica beads are attached to the ends of a membrane and then moved in opposite directions. The experimental data consist of force--displacement curves, which can be translated to stress--strain curves under certain assumptions. 

We conducted stretching, compression/buckling, and torsion simulations of \textsc{emb}s and \textsc{gv}s.
The stretching force--displacement curves of our model \textsc{emb}s and \textsc{gv}s were determined by adding oppositely equal forces to a small set of diametrically opposite vertices of their membrane. 
The \textsc{gv} torsion simulations were performed by rotating the \textsc{gv} cylinder ends in opposite directions, 
Fig.~\ref{fig: torsion_stress}. In principle, this could be achieved experimentally with an angular optical tweezer device, which is a relatively new methodology \cite{wang2007, wang2007b}.

We also derive analytical results for the stretching, torsion, and compression in the small deformation limit and verify their validity by simulations.

\subsubsection*{Microbubble stretching}
\addcontentsline{toc}{subsubsection}{Microbubble stretching}

\begin{figure}[!htbp]
\centering
\includegraphics[width=\textwidth]{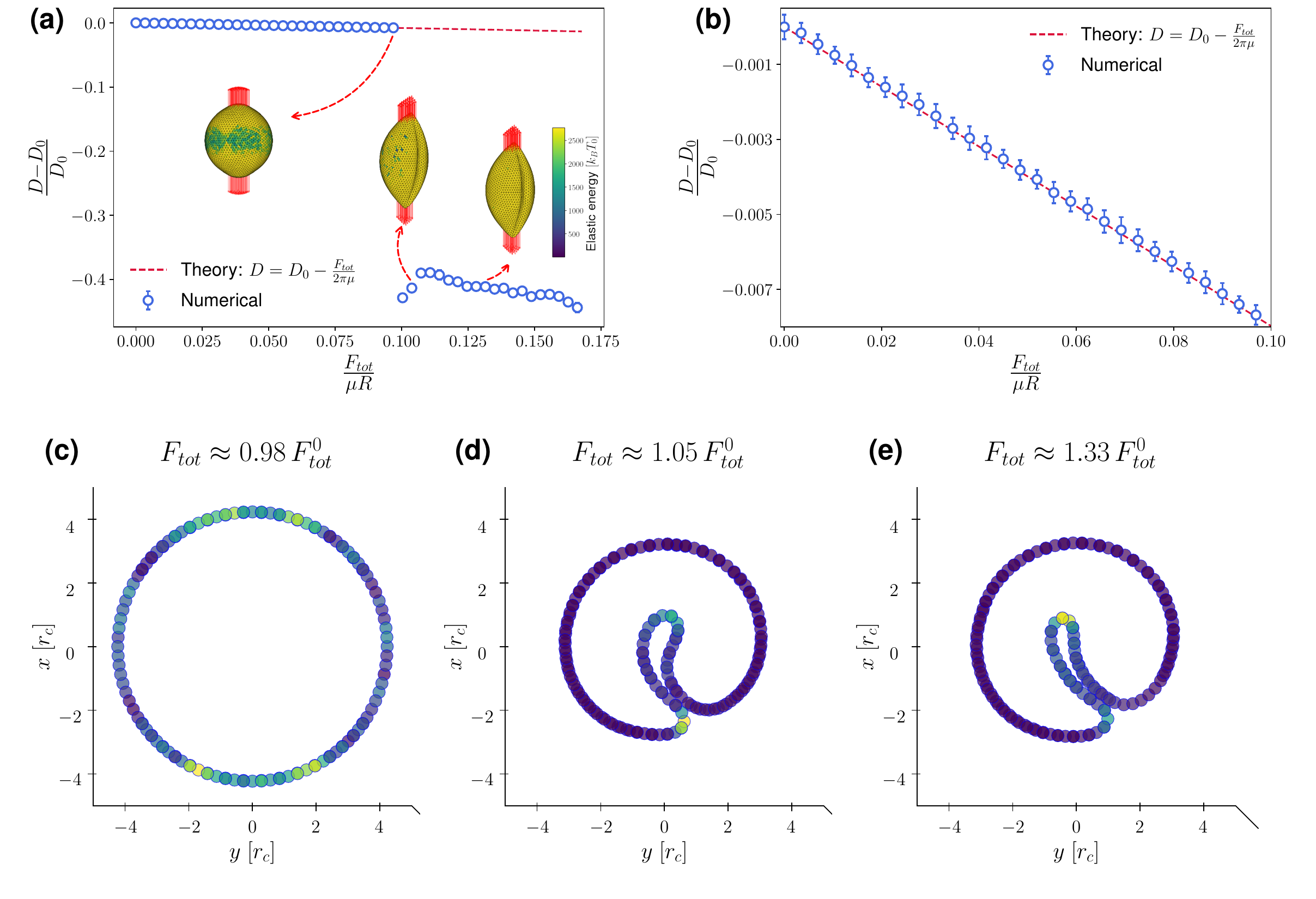}
\caption{(a) Relative \textsc{emb} diameter change as a function of applied force $F_{tot}$ in units of ${\mu R_{0}}$. (b) Small-strain part of the dependence in (a), compared to Eq.\,(\ref{emb_diam}). (c)-(e) Equatorial slices of \textsc{emb}s perpendicular to the stretching direction at values of $F_{tot}$, corresponding to the three regimes in (a); $F_{tot}^0 = 0.099 \mu R_{0}$.
The coloring in the side views (a) the top views (c)-(e) represents local elastic energy.
}
\label{fig: emb_diameters}
\end{figure}

\noindent
As shown in Fig.~\ref{fig: emb_diameters}(a), the stretching of an \textsc{emb} results in a decrease in diameter $D$ perpendicular to the stretching direction.
For small strains, one finds (Supplementary Information \ref{subsec:MBs-comp-buck})
\begin{equation}
\label{emb_diam}
    D = D_0 - \frac{F_{tot}}{2\pi \mu},
\end{equation}
where $F_{tot}$ is the total force applied to the ends of the \textsc{emb}, and $D_0=2R_0$ is its equilibrium diameter. 
As seen in Fig.~\ref{fig: emb_diameters}(a)(b), the measured diameter perpendicular to the stretching direction matches excellently with Eq.\,(\ref{emb_diam}) in the regime of small strains. 
The expression Eq.\,(\ref{emb_diam}) differs from the result for an elastic disc, which is often used to model red blood cell stretching in the linear regime \cite{henon1999}: $D = D_0 - \frac{F_{tot}}{2\pi \mu}\left[1-(1-\frac{\pi}{2})\frac{\mu}{K_a}\right]$.

Above $F_{tot} \approx 0.099\,\mu R_{0}$, the \textsc{emb} undergoes a mechanical instability into a circumferentially wrinkled shape and the diameter drops significantly.

\subsubsection*{Gas vesicle stretching}
\addcontentsline{toc}{subsubsection}{Gas vesicle stretching}

\begin{figure}[!htbp]
\centering
    \includegraphics[width=0.9\textwidth]{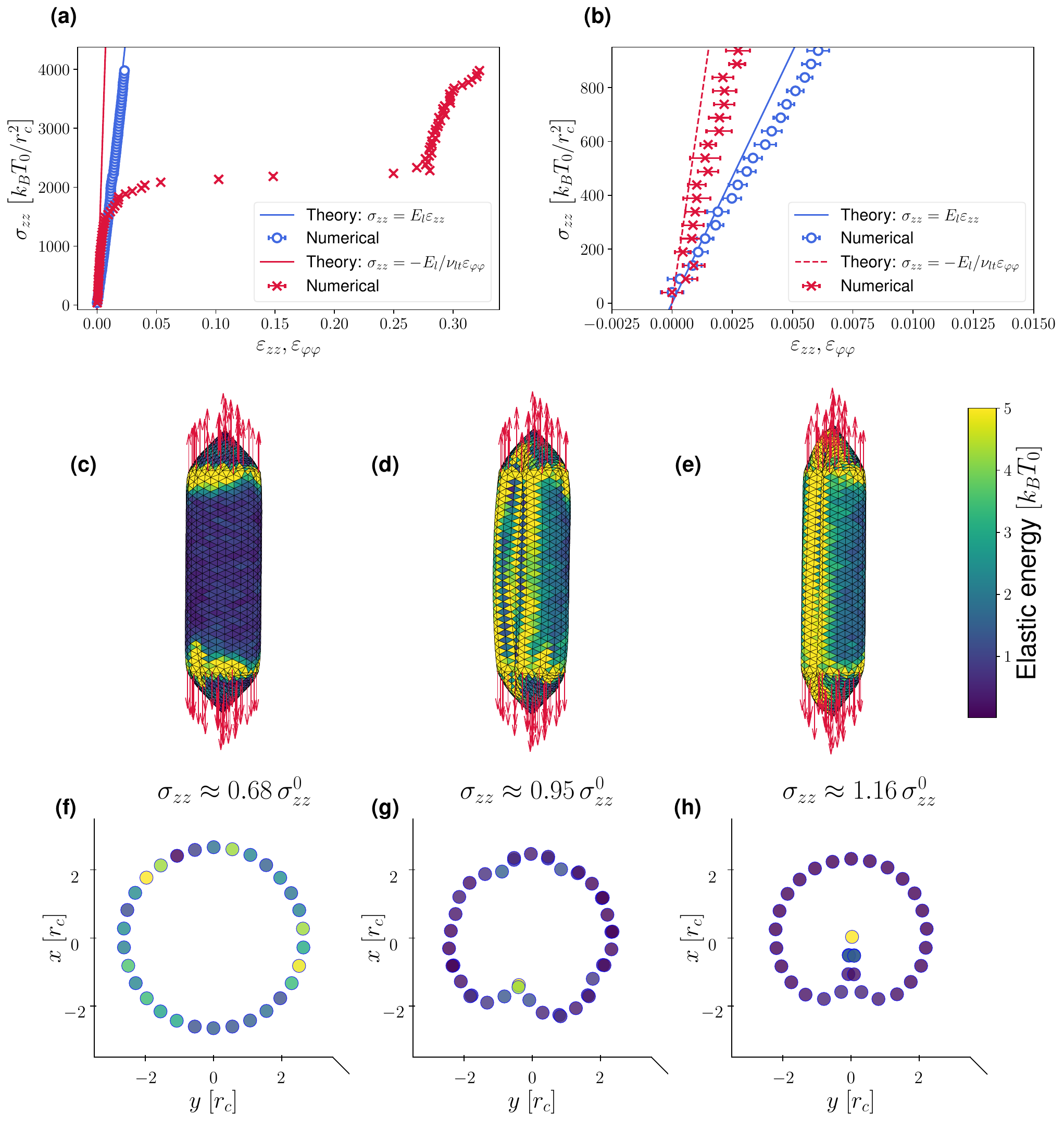}
\caption{(a) Stress $\sigma_{zz}$ as a function of longitudinal strain $\varepsilon_{zz}$ and circumferential strain $\varepsilon_{\varphi \varphi}$ for a \textsc{gv} stretched along its axis. (b) Comparison with Eqs.\,(\ref{gv-stretch1}) and (\ref{gv-stretch2}) of \ref{subsec:GVs-comp-buck} in the limit of low strain. (c)-(e) Side views and (f)-(h) central $xy$ slices of the \textsc{gv} at different values of $\sigma_{zz}$, with $\sigma_{zz}^0 = 2184\,k_BT_0/r_c^2$. 
The coloring represent local elastic energy.
To better depict the first buckling transition, slice (g) is taken at a height of $4\,r_c$ from the \textsc{gv} center. 
}
\label{fig: diameters}
\end{figure}

\noindent
We examine the response of a \textsc{gv} to stretching along its axis ($z$ axis), Fig.~\ref{fig: diameters}, which should be the most feasible experimentally.
For small stretching forces, the length of the \textsc{gv} increases linearly, while its diameter in the $xy$ plane decreases, in line with simple linear elastic response. 
The stress--strain response of a \textsc{gv} can be estimated by considering the uniaxial stress in its cylindrical part, $\sigma_{zz} = \frac{F_{tot}}{2\pi R}$, and the corresponding strain along $z$, $\varepsilon_{zz} = {(H_{cyl} - H_{cyl}^0)}/{H_{cyl}^0}$, where $H_{cyl}$ and $H_{cyl}^0$ are the heights of the \textsc{gv} cylinder in the deformed and reference configurations, respectively. 
The circumferential strain, on the other hand, is deduced from the relative radius change, $\varepsilon_{\varphi\varphi} = {(R - R_0)}/{R_0}$, where $R$ and $R_0$ are the radii in the deformed and reference configurations, respectively. 
It is important to emphasize that $H_{cyl}$ is measured between the ends of the cylindrical section, excluding the conical parts, where the stress depends on $z$ as the radius tapers from $R$ to $0$. 
Fig.~\ref{fig: diameters}(a)(b) shows that the linear response matches excellently with the constitutive equations Eqs.\,(\ref{gv-stretch1}) and (\ref{gv-stretch2}) of \ref{subsec:GVs-comp-buck}.
Measurement of this small-strain response could therefore be used to estimate both the longitudinal Young's modulus $E_l$ and the Poisson's ratio $\nu_{lt}$.

At a certain point, the \textsc{gv} buckles, forming a dimple, localized to each end of the cylindrical part, see Fig.~\ref{fig: diameters}(d)(g).
At a larger longitudinal stress $\sigma_{zz}^0 \approx 2184\,\frac{k_BT_0}{r_c^2}$, the \textsc{gv} buckles into a shape with a single dimple along the entire length, Fig.~\ref{fig: diameters}(e)(h). This transition is also accompanied by a large jump in the circumferential strain $\varepsilon_{\varphi \varphi}$. Interestingly, the longitudinal strain $\varepsilon_{zz}$ keeps increasing linearly with $\sigma_{zz}$ without a large discontinuity. 

\subsubsection*{Compression}\label{sec:compression}
\addcontentsline{toc}{subsubsection}{Compression}

\noindent
In compression experiments, external fluid pressure is increased and the corresponding volume change of the body is measured. In simulations, the pressure increase $\Delta p$ is controlled by varying the interaction parameter $a_{ww}$ (Eq.\,(\ref{cons})) between the water beads.

\begin{figure}[htbp]
    \centering
    \includegraphics[width=0.5\linewidth]{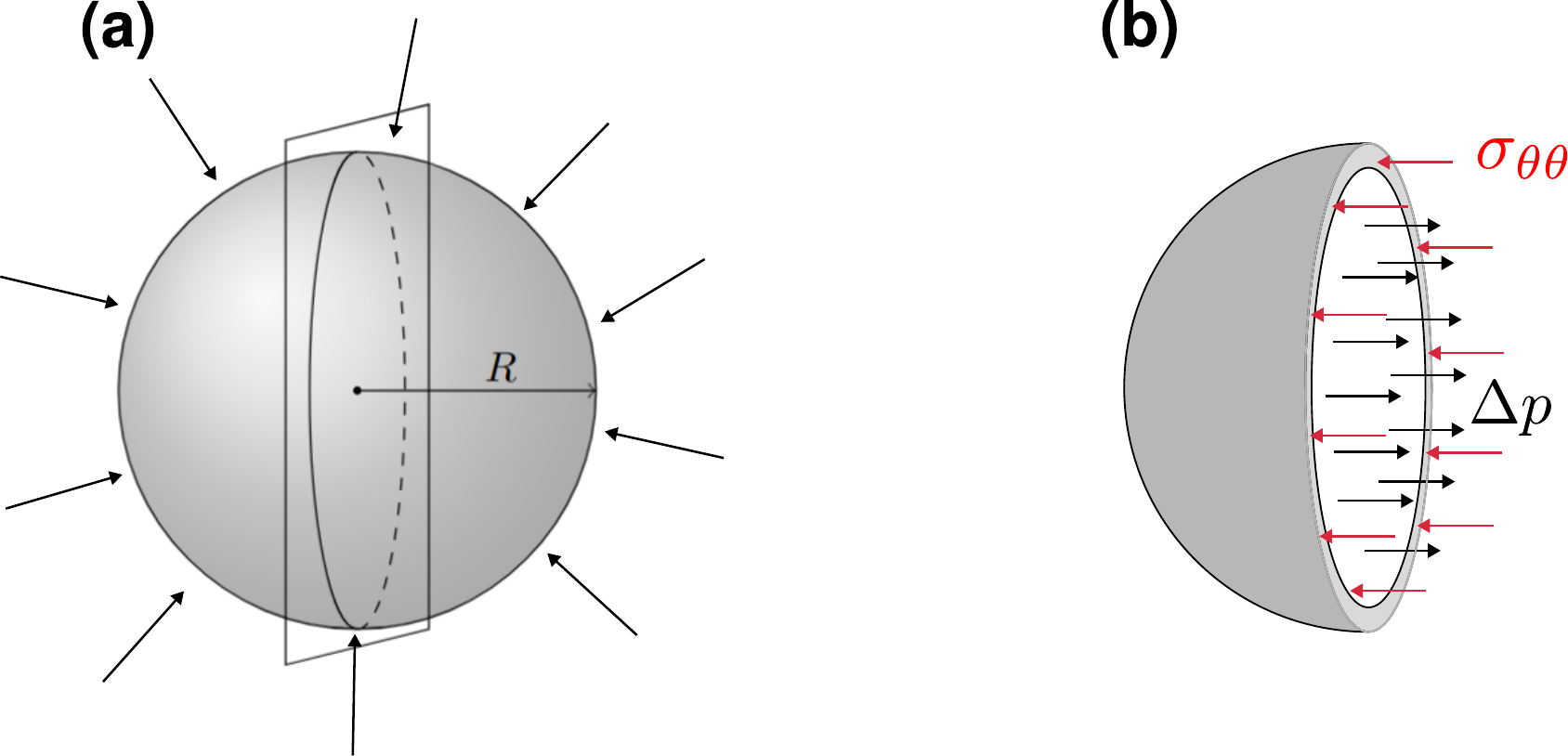}
    \caption{(a) Schematic representation of an \textsc{emb} in a pressurized environment, exerting uniform compression forces. (b) Differential pressure force on the hemisphere (black arrows) in equilibrium with elastic force of the shell (red arrows).}
    \label{fig: emb-press}
\end{figure}
\begin{figure}[!htbp]
    \centering
    \includegraphics[width=0.5\linewidth]{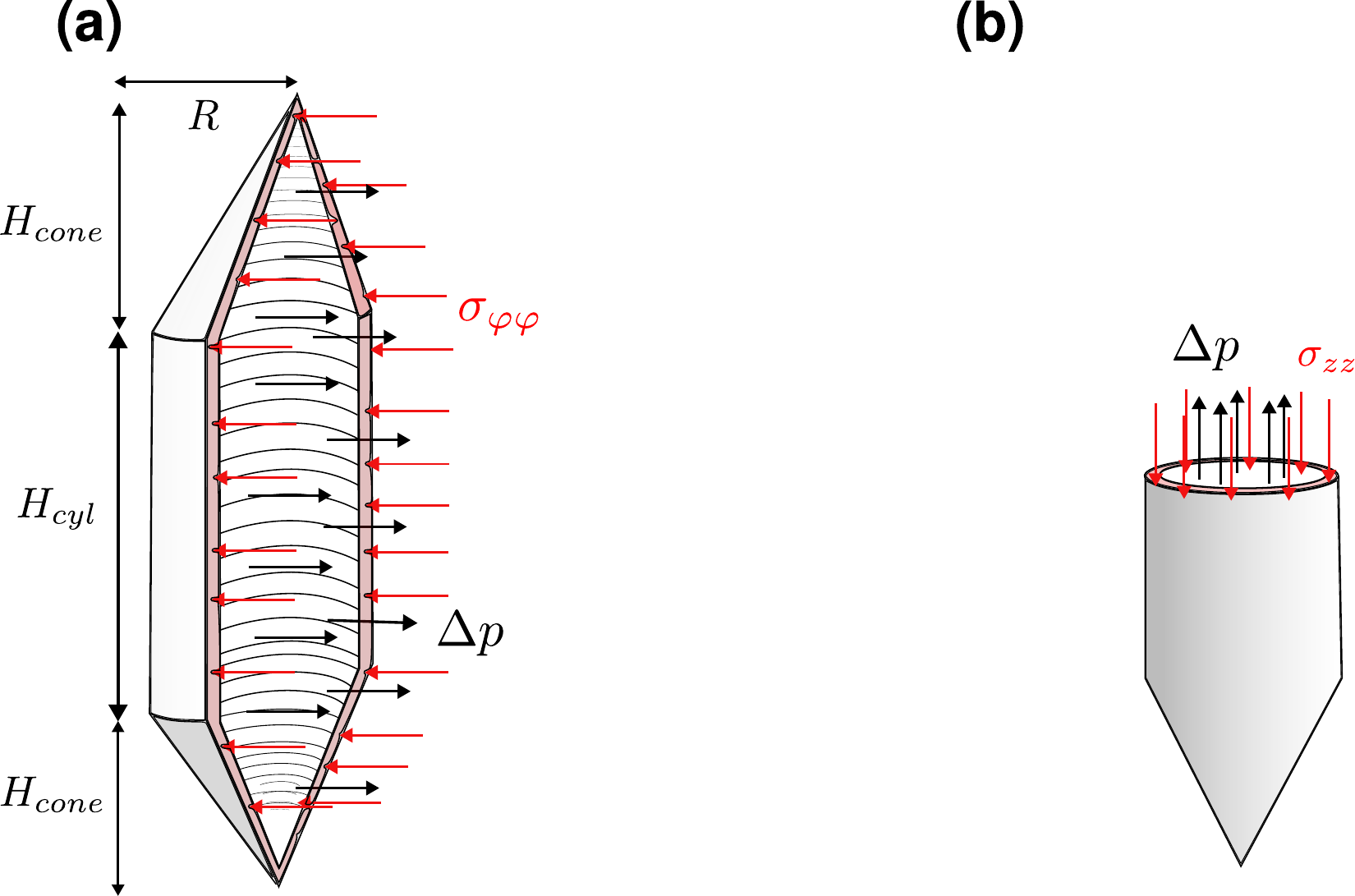}
    \caption{Schematic representation of a \textsc{gv} in a pressurized environment: (a) meridional stress $\sigma_{\varphi \varphi}$ caused by differential pressure on the longitudinal cross-section, and (b) equatorial stress $\sigma_{zz}$ caused by differential pressure on the transverse cross-section.}
    \label{fig: gv-press}
\end{figure}

For low $\Delta p$, the volume change can be obtained analytically for both \textsc{emb}s and \textsc{gv}s by calculating the stresses $\sigma_{\theta \theta}$ and $\sigma_{\varphi \varphi}$, depicted in Figs.~\ref{fig: emb-press} and \ref{fig: gv-press}.
Neglecting the compressibility modulus of the trapped gas and the membrane bending contributions, which are negligible in the regime of low deformation, the volume strain of an \textsc{emb} is (Eqs.\,(\ref{sigmathth})-(\ref{iso-comp}) of \ref{subsec:MBs-comp-buck})
\begin{equation}\label{vol_emb}
    \frac{\Delta V}{V_0} = - \frac{3(1-\nu)R_0}{2E}\Delta p = -\frac{3R_0}{4K_a}\Delta p,
\end{equation}
\begin{figure}[!htbp]
\centering
    \includegraphics[width=\textwidth]{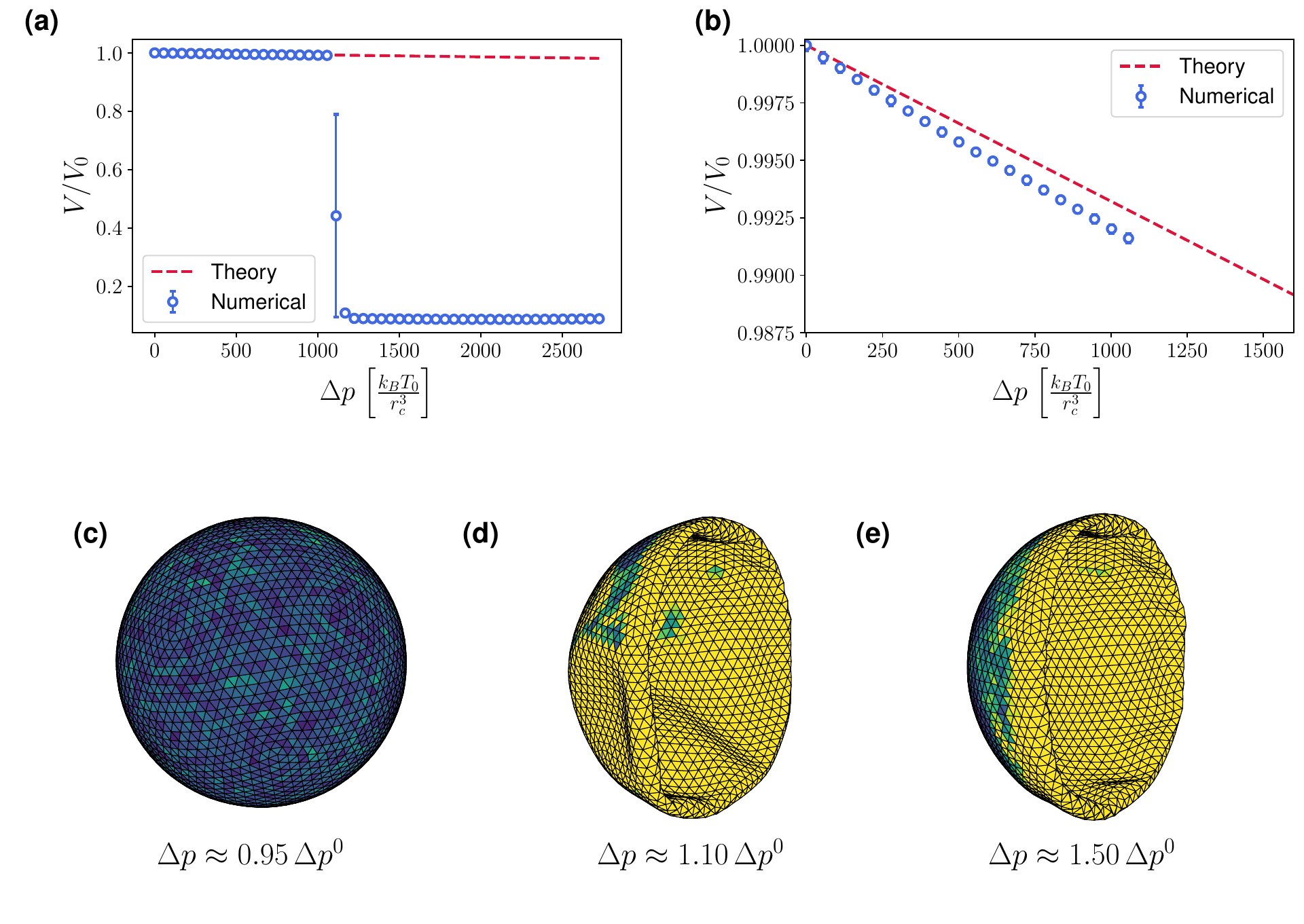}
\caption{(a) Relative volume change of an \textsc{emb} due to the external pressure increase $\Delta p$, compared to Eq.\,(\ref{vol_emb}) (red dashed line). (b) Closeup of (a) in the small deformation regime. (c)-(e) Compressed \textsc{emb} at different values of $\Delta p$, with $\Delta p^0 = 1113\,k_BT_0/r_c^3$ the buckling threshold pressure. Local buckling nucleations are noticeable in (c), while (d) and (e) show fully buckled states. The coloring represents local elastic energy.}
\label{fig: buckling_emb}
\end{figure}
which is confirmed numerically in Fig.~\ref{fig: buckling_emb}(a)(b). 

\begin{figure}[!htbp]
\centering
    \includegraphics[width=\textwidth]{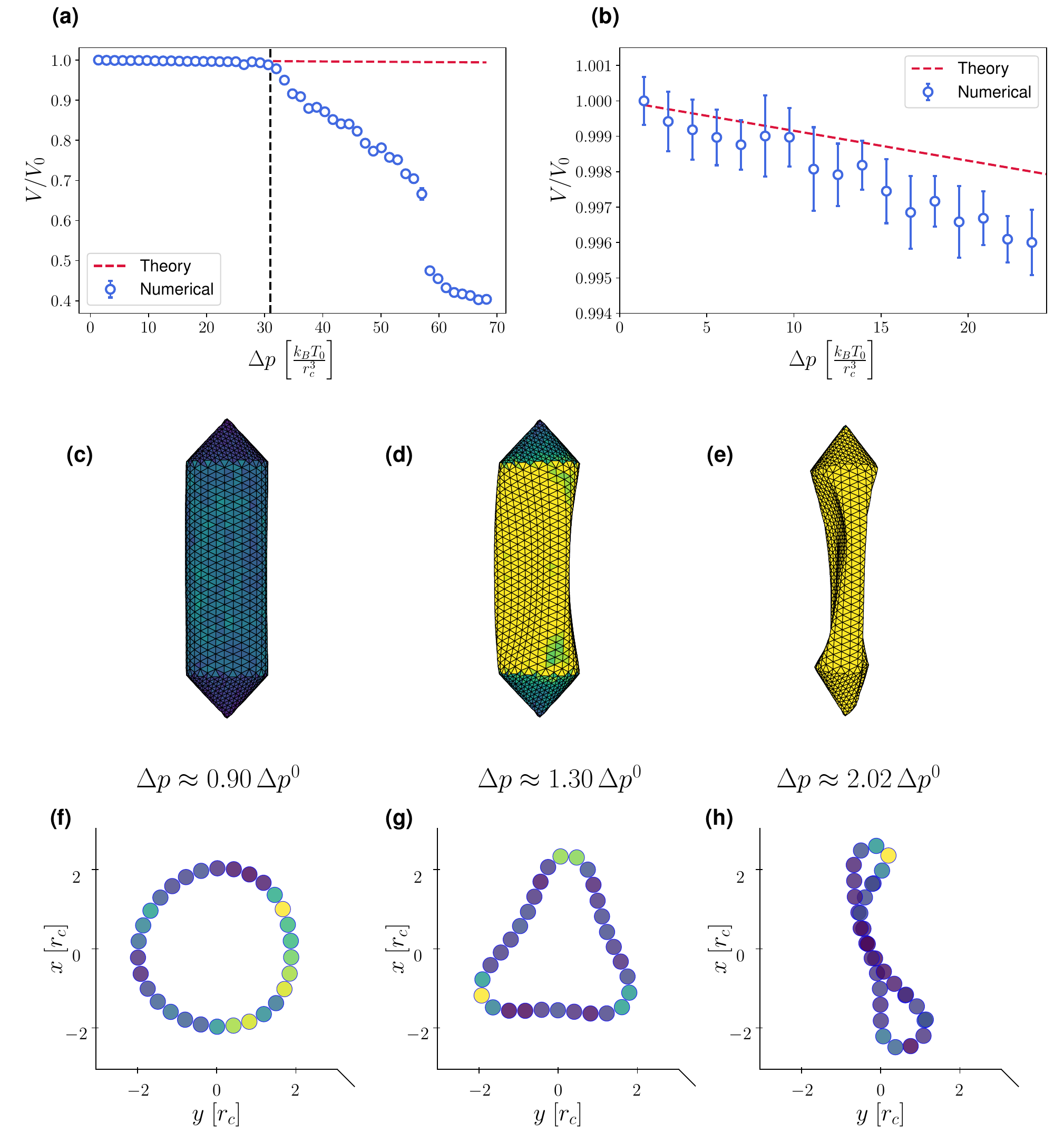}
\caption{(a) Relative volume change of a \textsc{gv} due to the external pressure increase $\Delta p$, compared to Eq.~(\ref{vol_gv}) (red dashed line). The buckling pressure $\Delta p^0 = 31.0\,k_BT_0/r_c^3$ is indicated by a black dashed vertical line. (b) Closeup of (a) in the small deformation regime. (c)-(e) Side views and (f)-(h) central transverse slices of the \textsc{gv} at different $\Delta p$. The coloring represents local elastic energy.}
\label{fig: compression}
\end{figure}

For a \textsc{gv} with orthotropic elasticity, the volume strain depends on both Young's moduli $E_l$ and $E_t$, and the Poisson's ratio $\nu_{lt}$ (Eqs.\,(\ref{epsilonzz})-(\ref{dVV0}) of \ref{subsec:GVs-comp-buck}):
\begin{equation}\label{vol_gv}
    \frac{\Delta V}{V_0} = - \frac{R_0}{2E_l}\left(1 - 4\nu_{lt} + 4\frac{E_l}{E_t}\right)\Delta p,
\end{equation}
which is again confirmed numerically in Fig.~\ref{fig: compression}(a)(b).
The linear weak-compression slope was also observed experimentally by Walsby \cite{walsby1982} using a glass compression tube and was used to estimate the bulk modulus of \textsc{gv}s. In the isotropic case, where $E_l = E_t$, Eq.\,(\ref{vol_gv}) agrees with the well-known result for isotropic cylindrical shells \cite{newman1957}, which is often used to estimate the Young's modulus of \textsc{gv}s \cite{walsby1982, walsby1994}.

In the small deformation regime governed by the membrane in-plane elasticity, as described by Eqs.\,(\ref{vol_emb}) and (\ref{vol_gv}), the effective compressibility of the shell increases linearly with its size $R_0$.
The shape of the shell also plays an important role. There is a distinct difference between the compression of the spherical \textsc{emb} and the cylindrical \textsc{gv} with the same radius.
For the same shell material, \textit{i.e.}, $E_l = E_t = E$, $\nu_{lt} = \nu$, Eq.\,(\ref{vol_gv}) gives 
$\frac{\Delta V}{V_0} = -\frac{(5-4\nu)R_0}{2E}\Delta p$, which should be compared with Eq.\,(\ref{vol_emb}).
Thus, for the same shell material and radius, the cylinder has almost twice the compressibility of the sphere.

\subsubsection*{Buckling}\label{sec:buckling}
\addcontentsline{toc}{subsubsection}{Buckling}
 
\textsc{gv}s are known to produce nonlinear acoustic signals when exposed to high-amplitude ultrasound pulses \cite{maresca2017}---at such pressure amplitudes, they buckle reversibly. By measuring optical density as a function of pressure, the buckling pressures were experimentally determined for a wide range of \textsc{gv} sizes and diameters. 
We investigate buckling instabilities of \textsc{emb}s and \textsc{gv}s by increasing the solvent pressure beyond the small strain limit.

Examples of \textsc{emb} buckling are shown in Fig.~\ref{fig: buckling_emb}(c)-(e).
An analytical expression for the critical buckling pressure of spherical shells exists, Eq.\,(\ref{eq: pcrit}) in \ref{subsec:MBs-comp-buck}, which is in an excellent agreement with our simulations, see Fig.~\ref{fig: pcrit} in \ref{subsec:MBs-comp-buck}. It involves an empirical correction factor due to imperfections of the shell, which consistently takes on a well-defined value also for our model \textsc{emb}. 
We also performed dynamic numerical buckling experiments at various pressure increase rates. While the buckling pressure remains nearly unchanged across different rates, the structure evolves through different intermediate shapes, from a single indentation for the slowest pressure increase to multiple lobes when the pressure increases more abruptly, which is in line with the numerical simulations of Ref.~\cite{gompper2011}.

For \textsc{gv}s, we compare the predicted buckling pressure to the hydrostatic critical pressure in Ref.~\cite{cherin2017}, which considers isolated \textit{Halobacterium salinarum} \textsc{gv}s in a solution. 
At a pressure of about $\Delta p^0 \approx 31.0\,k_BT_0/r_c^3$, our \textsc{gv} buckles into a shape with three lobes along the circumference, Fig.~\ref{fig: compression}(d)(g). Reverting to physical units and up-scaling by $f_{scale}$ (see Methods and Table \ref{tab:scales} in \ref{sec:compdetails}), $\Delta p^0 \approx 37.9\,$kPa. This is below the value of 64\,kPa in Ref.~\cite{cherin2017}. It is also significantly below experimental values of Refs.~\cite{lakshmanan2016} and \cite{zhang2020vibration} of about $\Delta p^0 \approx 200\,$kPa and $\Delta p^0 \approx 178\,$kPa, respectively. 
One possible reason for the discrepancies with Refs.~\cite{lakshmanan2016, zhang2020vibration} is that these experiments used acoustic waves to induce the \textsc{gv} collapse, which is known to yield higher measured collapse pressures than hydrostatic techniques. 
A general reason is that the actual bending constant $\kappa$ is an independent parameter, lower than predicted by thin-shell theory---a single molecular layer membrane can hardly be considered a continuum across its thickness. 
Optimal values of the elastic coefficients and their uncertainties can be estimated directly from experimental measurements using hierarchical Bayesian uncertainty quantification \cite{economides2021}, which will be addressed in our future work. 

At an even higher pressure of $\Delta p \approx 58\,k_BT_0/r_c^3 \approx 71\,$kPa, the \textsc{gv} undergoes a transition into a two-lobe shape, Fig.~\ref{fig: compression}(e)(h), which is actually closer to the collapse pressure in Ref.~\cite{cherin2017}. This particular shape was also obtained via a linear buckling analysis by Salahshoor \textit{et al.} \cite{shapiro2022}. 
Both buckling shapes are corroborated by the existence of two low-frequency vibrational modes of similar shape, which are calculated in the next section.

\subsubsection*{Vibrational modes}
\addcontentsline{toc}{subsubsection}{Vibrational modes}

\noindent
Finding the right ultrasound frequency is key for optimal tissue imaging \cite{Bouakaz:2003}, as well as for inducing sonophoresis and affecting drug carrier behavior in terms of growth, oscillations, rupture, and drug release via cavitation \cite{Zhang:2016}. Here, we study the low-frequency modes of \textsc{gv}s that could play a major role in acoustic backscattering in ultrasound experiments.

\begin{figure}[!htbp]
    \centering
    \includegraphics[width=\linewidth]{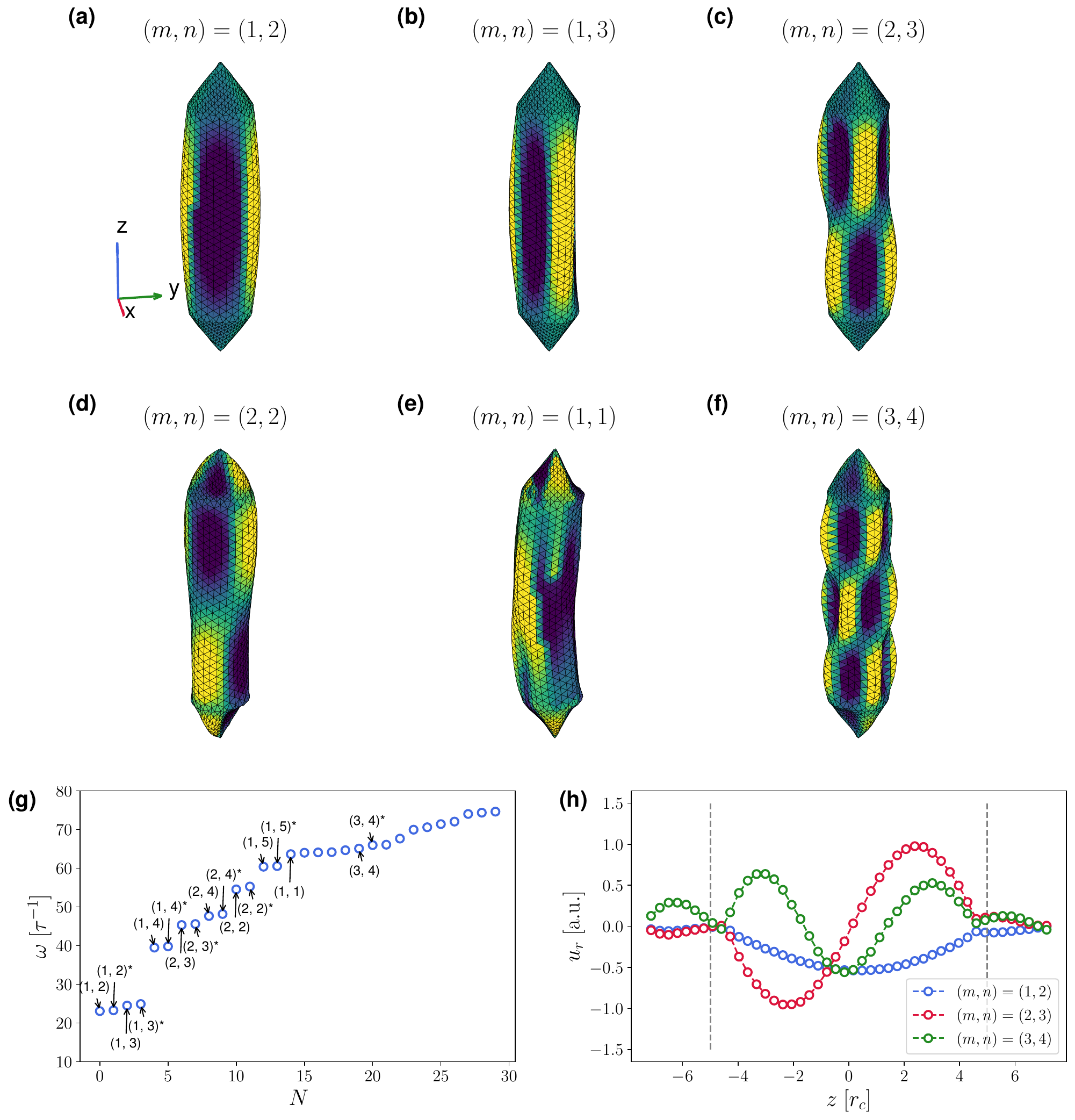}
    \caption{(a)-(f) Selected low-frequency eigenmodes of a \textsc{gv} with corresponding axial and circumferential numbers $m$, $n$, ordered by increasing frequency. The mesh triangles are colored according to modal displacement. (g) First 30 eigenfrequencies with selected modes labeled by their mode indices $(m, n)$. The star $(m, n)^*$ denotes an eigenmode rotated by $\frac{\pi}{2n}$ around the $z$ axis, relative to the mode with the same indices $(m, n)$. (h) The $u_r(z)$ profile (axial profile) of selected modes. The dashed lines indicate the ends of the \textsc{gv} cylinder.}
    \label{fig: eig}
\end{figure}

To extract the eigenmodes and their corresponding eigenfrequencies, we run a long simulation of $4000\,\tau$ (Eq.\,(\ref{tau}), Table \ref{tab:scales}) of a \textsc{gv} surrounded by \textsc{dpd} water and perform principal component analysis (PCA) on the trajectories of the \textsc{gv} vertex beads, see Methods. 
Some of the extracted low-frequency modes are shown in Fig.~\ref{fig: eig}(a)-(f).

In the relevant regime, where the \textsc{gv} cylinder is longer than the end cones, the lowest-frequency modes are largely confined to the cylindrical region, as highlighted by Fig.~\ref{fig: eig}(h). Consequently, the ends of the cylinder impose an effectively rigid boundary condition.
In this regime, 
the modes can be classified not only according to their circumferential number $n$, but also their axial number $m$, on top of their polarization branch, Fig.~\ref{fig: eig}(a)-(f).
They follow considerably well the solutions to the linearized equations (Donnel theory \cite{donnell1934}) of simply supported cylindrical shell (shear diaphragm at both ends at $z=-\textstyle{\frac12} H_{cyl}^0$ and $z=\textstyle{\frac12} H_{cyl}^0$) \cite{amabili2018}, ${\bf u}(z, \varphi, t) = \left(u_r \hat{\bf e}_r + u_\varphi \hat{\bf e}_\varphi + u_z \hat{\bf e}_z\right)\cos(\omega t)$, with
\begin{align}
    u_r &= C_r \sin(\lambda_m z + \textstyle{\frac{m\pi}{2}})\cos (n\varphi) \label{ur}\\
    u_\varphi &= C_\varphi \sin(\lambda_m z + \textstyle{\frac{m\pi}{2}}) \sin(n\varphi) \label{uphi} \\
    u_z &= C_z \cos(\lambda_m z + \textstyle{\frac{m\pi}{2}}) \cos(n\varphi), \label{uz} 
\end{align}
where $u_r$, $u_\varphi$, $u_z$ are displacement fields of the cylindrical shell in $r$, $\varphi$, and $z$ directions, $\lambda_m = \frac{m\pi}{H_{cyl}^0}$, and $C_r$, $C_\varphi$, $C_z$ are the amplitudes, which depend on $n$, $m$, and the elastic coefficients \cite{amabili2018}. 
For every mode with $n\ne 0$, there exists another mode with the same frequency, rotated by $\frac{\pi}{2n}$ around the $z$ axis, i.e.,  with $\cos(n\varphi)$ and $\sin(n\varphi)$ in Eqs.~(\ref{ur})-(\ref{uz}) interchanged.

The lowest-frequency mode is the doubly degenerate $(m,n) = (1,2)$ mode, Fig.~\ref{fig: eig}(a), with $\omega \approx 23.1\,\tau^{-1}$, corresponding to frequency $\nu \approx 201\,$MHz. It resembles the \textsc{gv} shape at the second buckling transition, Fig.~\ref{fig: compression}(e)(h).

Our results can be compared with the FEM simulations of \textsc{gv}s in Ref.~\cite{shapiro2022}, where the lowest-frequency vibrational mode is the $(m,n)=(1,1)$ mode at $\nu = 328\,$MHz. The lower frequency in our case is due to the larger diameter of \textsc{gv}, $140$\,nm, compared to $85$\,nm in Ref.~\cite{shapiro2022}. Interestingly, in our case, the $(m,n)=(1,1)$ mode appears only as the 15th mode in Fig.~\ref{fig: eig}(g), with a frequency of $\nu \approx 555\,$MHz.

\subsection*{Rheological properties}
\addcontentsline{toc}{subsection}{Rheological properties}

\subsubsection*{Behavior in shear flow}
\addcontentsline{toc}{subsubsection}{Behavior in shear flow}

\noindent 
Considering the effects of shear forces on \textsc{gv} dynamics is crucial for their use in targeted drug delivery within the bloodstream.
Since \textsc{gv}s are elongated and exhibit a wide range of aspect ratios, their behavior in shear flow can be compared to that of ellipsoids. In shear flow, neutrally buoyant ellipsoids undergo rotations known as Jeffery orbits \cite{Jeffery:1922}. During these rotations, the ellipsoid is continuously flipping in the shear plane (the plane spanned by the velocity and its gradient), and the angle $\theta$ between its axis and the velocity evolves as
\begin{equation}
    \tan(\theta) = \frac{a}{b} \tan\left(\frac{ab \dot{\gamma}\,t}{a^2 + b^2}\right),
    \label{eq:jeffery}
\end{equation}
where $a$ and $b$ are the major and minor axes of the ellipsoid and $\dot{\gamma}$ is the shear rate.

\begin{figure}[!htbp]
\centering
\includegraphics[width=\textwidth]{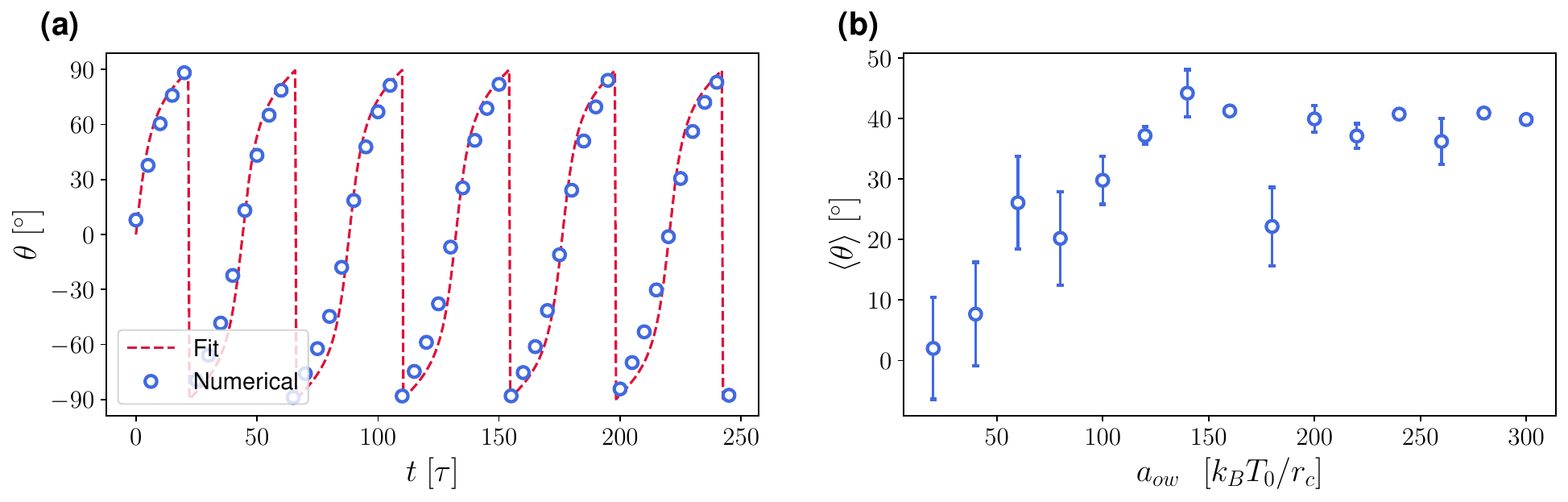}
\caption{(a) Time dependence of the \textsc{gv} axis inclination angle $\theta$ with respect to the flow direction for \textsc{gv}--water repulsion strength $a_{ow} = 50\,k_{B}T_0/r_{c}$ and \textsc{gv}--water dissipative interaction strength $\gamma_{ow} = 40\,m/\tau$ for $\dot{\gamma} = 0.0862\,\tau^{-1}$. Numerical values (blue circles) are compared to Eq.\,(\ref{eq:jeffery}) (red dashed line). (b) Dependence of mean $\theta$ on $a_{ow}$ for $\dot{\gamma} = 0.0862\,\tau^{-1}$.
}
\label{fig: shear_all}
\end{figure}

Using open-boundary molecular dynamics (\textsc{obmd}) (see Methods), we generate shear flow and apply it to a \textsc{gv} fixed at the center of the domain but free to rotate.
In our model, the dissipative interaction between the \textsc{gv} vertex beads and water beads, taking into account Eq.\,(\ref{gamma_owg}), ensures no-slip boundary condition. This boundary condition leads to a tumbling motion consistent with Eq.\,(\ref{eq:jeffery}), as shown in Fig.~\ref{fig: shear_all}(a). In principle, one could functionalize \textsc{gv}s with a hydrophobic surfactant. In such a case, the no-slip boundary condition may no longer apply, potentially altering the qualitative behavior of \textsc{gv}s in shear flow. To investigate this, we explore the \textsc{gv} response in shear flow depending on the repulsive interaction strength $a_{ow}$ between the vertices and water beads.

The dependence of the mean inclination angle $\langle \theta\rangle$ on $a_{ow}$ is shown in Fig.~\ref{fig: shear_all}(b). For $0<a_{ow}<50\,k_{B}T_0/r_{c}$, the \textsc{gv} exhibits tumbling behavior. Above $a_{ow}\approx 50\,k_{B}T_0/r_{c}$, the \textsc{gv} aligns at a fixed angle in the shear plane, which increases to approximately $40^\circ$ in the limit of large $a_{ow}$. Here, $\gamma_{ow}$ was fixed to the value required by Eq.\,(\ref{gamma_owg}), which guarantees no-slip boundary conditions for $a_{ow}=0$.

\section*{Discussion}
\addcontentsline{toc}{section}{Discussion}

The growing demand for encapsulated materials in theranostic ultrasound applications underscores the need for versatile and robust modeling approaches. Current theoretical frameworks often rely on oversimplified descriptions, and while fully atomistic models provide detailed insights, they are computationally prohibitive for large-scale or dynamic simulations. In this paper, we developed a general framework for modeling shelled biomaterials, and presented particle-based mesoscopic models of \textsc{emb}s and \textsc{gv}s as two representative applications. 
The elastic energy of their membrane was derived from the continuum theory of elasticity and discretized on a triangular surface, where we drew inspiration from \textsc{rbc} membrane network models. The framework can be easily applied to anisotropic membranes, such as \textsc{gv}s, which exhibit orthotropic elasticity due to their increased stiffness along the ridges of the GvpA protein. 

Our description of membrane elasticity builds on the same theoretical foundation as existing force fields, encompassing previous models, e.g. for \textsc{rbc}s \cite{Svetina:1982, Svetina:1983, Svetina:1989, Svetina:1998, Heinrich:1999, Derganc:2003, Ziherl:2008, Fedosov:2010, Bian2020, economides2021} or viral capsid shells \cite{podgornik2012, podgornik2020}, while further extending them by incorporating anisotropic elastic terms, here applied to \textsc{gv}s.
The elasticity of computational shell models is governed by elastic coefficients, including Young’s moduli, Poisson’s ratios, bending constants, as well as coefficients of mutual influence, which describe the coupling between extensional and shear strains, and Chentsov coefficients, which characterize the interaction of shear strains across different planes. These coefficients can be determined experimentally. Consequently, any membrane composition can be modeled by our framework, as long as its elastic properties are known. This has important implications for fields such as bioengineering and medical applications where elastic properties are concerned, as our framework can model a wide range of membrane-based systems of arbitrary shapes and local anisotropy, from biological cells to artificial capsules.

We validated the framework by simulating stretching, buckling, and shear flow dynamics of \textsc{emb}s and \textsc{gv}s, comparing our results with theoretical predictions in the limit of small deformations. The stress-strain curves obtained from the stretching experiments agree well with the theory of linear deformation for both \textsc{emb}s and \textsc{gv}s. Furthermore, we were able to reproduce the relationship between the critical buckling pressure and the membrane radius. We also derived analytical expressions that can be used for the experimental determination of the elastic coefficients.

However, the mechanical properties of fluid-immersed objects depend not only on the material of the membrane but also on their environment and the interactions with it.
These interactions significantly influence the rheological behavior, which is crucial when modeling membranes in flow. 
\textsc{dpd} allows tuning of fluid--object interactions, thereby influencing the dynamic coupling between objects and the fluid.
Proper adjustment of the dissipative parameter $\gamma_{ow}$ enables control over the degree of slip in the boundary condition at the membrane--solvent interface, ranging from no-slip to full slip.
Our shear flow numerical experiments demonstrated the rotational periodic motion (flow tumbling) of a single \textsc{gv} suspended in a \textsc{dpd} solvent, consistent with Jeffery's theoretical predictions. 
Moreover, we find that both the repulsion and dissipative parameters, $a_{ow}$ and $\gamma_{ow}$, play a crucial role in \textsc{gv} motion, either inducing flow tumbling or leading to flow alignment at a specific inclination angle.

While continuum models focus on modeling the dynamics of individual bubbles to elucidate fundamental principles \cite{dollet2019}, our \textsc{dpd} approach extends beyond these limitations by allowing the simulation of multiple interacting bodies, thus capturing collective behavior and complex interactions that are critical in many applications.  
The presented modeling framework is a first step towards large-scale simulations of multiple micro- and nanostructures and their interaction with propagating ultrasound waves. Modeling the ultrasound propagation on mesoscopic scales with our virtual \textsc{us} machine \cite{masa2025, petra2022, noguchi2020, noguchi2022} will assist and advance simulation-driven optimization of ultrasound-based theranostics. 
The optimal parameters that reproduce the given experiments can be efficiently determined using Bayesian uncertainty quantification \cite{economides2021, lucas2023}. The proposed computational framework would allow for controlled testing, data-driven quantification of uncertainties, and rational optimization of experimental \textsc{us} parameters. 

\section*{Methods}
\addcontentsline{toc}{section}{Methods}
The simulations were carried out using Mirheo \cite{alexeev2020}, a high-throughput simulation package, specifically designed and optimized for \textsc{dpd} simulations. While isotropic elastic forces are already implemented in Mirheo, we modified and extended its functionalities to include orthotropic elastic forces, the gas pressure contribution as well as the \textsc{obmd} used in nonequilibrium simulations.

\subsection*{Dissipative particle dynamics}
\addcontentsline{toc}{subsection}{DPD}
\label{sec:dpd}

The \textsc{dpd} method \cite{Hoogerbrugge:1992} is a particle-based mesoscopic simulation technique that allows modeling of fluids and soft matter. A \textsc{dpd} system is represented by $N$ particles, which interact through pairwise effective potentials and move according to the second law of Newton. In a \textsc{dpd} simulation, a particle represents a cluster of molecules, and the position and momentum of the particle are updated constantly, but distributed at discrete time steps. We use \textsc{dpd} to represent large numbers of water or gas molecules as single beads.

The interparticle force $\mathbf{F}_{ij} = \mathbf{F}_{ij}^C + \mathbf{F}_{ij}^D + \mathbf{F}_{ij}^R$ exerted by bead $j$ on bead $i$ consists of conservative, dissipative, and random forces \cite{espanol1995}
\begin{align}
    \mathbf{F}_{ij}^C &= a_{\alpha\beta} \omega_C(r_{ij})\uvec{r}_{ij} \label{cons}\\
    \mathbf{F}_{ij}^D &= - \gamma_{\alpha\beta} \omega_D(r_{ij})\left(\vec{v}_{ij}\cdot\uvec{r}_{ij}\right)\uvec{r}_{ij}\label{diss} \\
    \mathbf{F}_{ij}^R &= \sigma_{\alpha\beta} \omega_R(r_{ij})\Theta_{ij}\uvec{r}_{ij},\label{random}
\end{align}
where $a_{\alpha\beta}$ and $\gamma_{\alpha\beta}$ represent the conservative and dissipative parameters for a bead pair of species $\alpha$ and $\beta$ (water, gas, \textsc{uca} object), specified in Table \ref{tab:dpd} of \ref{sec:dpdinter}, 
$\sigma_{\alpha \beta}$ is the random force amplitude connected with the dissipative parameter $\gamma_{\alpha\beta}$ (\ref{sec:weights}), $\uvec{r}_{ij}=\frac{\vec{r}_i-\vec{r}_j}{|\vec{r}_i-\vec{r}_j|}$ is the normalized vector along the inter-particle axis, $\vec{v}_{ij}=\vec{v}_i-\vec{v}_j$ the relative velocity of the two interacting particles, $\omega_C, \omega_D$, $\omega_R$ are the weight kernels described in \ref{sec:weights}, while $\Theta_{ij}=\Theta_{ji}$ is a zero-mean random Gaussian variable with unit variance, uncorrelated between different pairs of particles $i$ and $j$,
\begin{equation}
    \langle \Theta_{ij}(t) \Theta_{kl}(t') \rangle = (\delta_{ik}\delta_{jl} + \delta_{jk}\delta_{il})\delta(t-t').
\end{equation}

The 
equation of state of a \textsc{dpd} fluid is quadratic in the density,
\begin{equation}\label{peos}
    p = \rho k_BT_0 + \alpha a_{ww} \rho^2,
\end{equation}
where $\alpha \approx 0.100\,r_c^4$ is a semi-empirical prefactor \cite{groot1997}.

\noindent
\subsubsection*{Fundamental scales}
\addcontentsline{toc}{subsubsection}{Fundamental scales}
\label{sec:scales}

\noindent
The solvent and encapsulated gas phases are modeled using \textsc{dpd}, which allows capturing relevant rheological properties, such as viscosity, on large spatiotemporal scales \cite{marsh1997}. To parameterize the interactions between the various \textsc{dpd} beads, we first select the appropriate coarse-graining level, or equivalently, the length scale $r_c$. Due to the disparate sizes of \textsc{emb}s, which can be as large as several micrometers, and \textsc{gv}s, where the diameter is at most several hundred nanometers, we use two different sets of fundamental scales (length, energy, and mass), specified in Table~\ref{tab:scales} of Supplementary Information \ref{sec:compdetails}. 

We choose the length scale $r_c$ so that the radius of the particular object in the smallest dimension is at least $2\,r_c$. This is to ensure a large enough resolution of the immersed objects compared to the cutoff $1\,r_c$ of the \textsc{dpd} interaction between the beads, Eqs.\,(\ref{omegaC})-(\ref{eq_power}) of \ref{sec:weights}. The mass scale $m$ is chosen to reproduce the density of the water/gas/shell and is calibrated, so that the mass of the water bead is equal to $1\,m$: $m = {\rho_w^{exp} r_c^3/\rho_w}$, where $\rho_w$ is the number density of water beads and $\rho_w^{exp} = 997\,$kg/m$^3$ is the physical density of water. For the energy scale $\varepsilon$, we take the thermal energy at room temperature $T_0 = 300$\,K, $\varepsilon = k_BT_0$. The time scale $\tau$ follows from the other three fundamental scales:
\begin{equation}
    \tau = \sqrt{\frac{m r_c^2}{\varepsilon}}.
    \label{tau}
\end{equation}

We use the \textsc{dpd} parameters specified in Table \ref{tab:dpd} of Supplementary Information \ref{sec:compdetails}. To keep the \textsc{dpd} system fluid-like and avoid the freezing artifacts appearing at $a_{\alpha\beta} \gtrsim 250$\,${{k_BT_0}/ {r_c}}$ \cite{Dzwinel:2000,Trofimov:2003,pivkin2006}, we use $a_{\alpha\beta}=100$\,${{k_BT_0}/{r_c}}$. For high coarse-graining levels, as is the case here, such a value of $a_{\alpha\beta}$ leads to a highly compressible liquid or equivalently a low value of the speed of sound $c_0$. Since we are interested in phenomena characterized by a low Mach number $\mathrm{Ma}={u}/{c_0} \ll 1$, with $u$ the typical particle velocity, this does not have a great impact on dynamics in this work. 
We have also ensured that the water/gas viscosity ratio is high by varying $\gamma_{gg}$ and the kernel power $k_{gg}$ in Eq.\,(\ref{eq_power}) of \ref{sec:weights}. 
Our choice of \textsc{dpd} parameters yields realistic viscosity ratios $\eta_w/\eta_{N_2} \approx 63$ for water/nitrogen and $\eta_w/\eta_{air} \approx 48$ for water/air, which are close to the experimental values of $\eta_w/\eta_{N_2} \approx 51$ and $\eta_w/\eta_{air} \approx 48$.

\subsubsection*{Down-scaling of elastic forces}
\addcontentsline{toc}{subsubsection}{Down-scaling of elastic forces}
\label{sec:reduced}

\noindent
At typical length scales $r_c$ of the objects, the dimensionless values of the 2D Young's moduli are large, as they scale as $\sim r_c^2/\varepsilon$, \textit{i.e.}, physically, elastic energy is much larger than $k_BT_0$.
To ensure computational feasibility, we scale down all elastic moduli in our simulations by a factor $f_{scale} \ll 1$. 
This preserves the so-called Föppl-von-Karman number $FvK$, which determines the shape of the objects in equilibrium \cite{Paulose:2012,economides2021},
\begin{equation}
FvK =  \frac{E R_{0}^2}{\kappa},   \label{foppl}
\end{equation}
where $E$ is the 2D Young's modulus, $\kappa\propto E$ is the bending constant, and $R_{0}$ is the typical radius of the object. 
Consequently, computed quantities such as critical stretching forces and buckling pressures must be multiplied by $f_{scale}$.

The behavior of elastic objects under shear flow is governed by the dimensionless capillary number
\cite{Gennes:1985,economides2021}
\begin{equation}
  Ca = \frac{\eta\dot{\gamma}R_{eff}}{\mu}  \label{eq:cap}.
\end{equation}
To preserve it, the viscosity of water must also be scaled accordingly, $\tilde \eta_w = f_{scale} \eta_w$, where $\eta_w$ is the physical viscosity of water. 

We choose two different $f_{scale}$ values corresponding to each unit set (Table \ref{tab:scales} in Supplementary Information \ref{sec:compdetails}), ensuring that ${\kappa}/(k_BT_0) > 10$ to prevent significant perturbation of the object by thermal fluctuations, which primarily originate from the solvent beads.
For each unit set, the target viscosity $\tilde \eta$ is achieved by adjusting the values of $\gamma_{ww}$ and $k_{ww}$ (Table \ref{tab:dpd} in Supplementary Information \ref{sec:dpdinter}).

\subsubsection*{Fluid--structure interactions}
\addcontentsline{toc}{subsubsection}{Fluid-structure interactions}
\label{sec:fsi}

\noindent
The boundary conditions at the fluid-immersed structure interface have a large effect on the behavior of \textsc{uca}s under non-equilibrium conditions, such as shear or plug flow \cite{Tsuji:1992}. Firstly, to prevent leakage of the solvent inside the gas vesicles or microbubble, we impose the no-through boundary condition by using the bounce-back mechanism, where the particles are introduced back based on the Maxwell distribution of velocities at temperature $T_0$ \cite{revenga1999}. 

To control the velocity boundary conditions at the \textsc{uca}--water or \textsc{uca}--gas interface, one typically tunes the dissipative parameter $\gamma_{ow}$ ($\gamma_{og}$) between the water (gas) and the \textsc{uca} beads. A specific value of the dissipative parameter ensures the no-slip boundary condition \cite{fedosov2010}:
\begin{equation}
    \gamma_{o\{w,g\}} = \frac{2\tilde \eta_{\{w,g\}} (2k_{fsi}+1)(2k_{fsi}+2)(2k_{fsi}+3)(2k_{fsi}+4)}{3\pi r_c^4 \rho_{\{w,g\}} \rho_m},
    \label{gamma_owg}
\end{equation}
where $\rho_m$ is the area number density of the vertices of the immersed membrane, $\rho_{\{w,g\}}$ is the number density of the fluid (water or gas), $\tilde \eta_{\{w,g\}}$ is the corresponding scaled-down viscosity, and $k_{fsi}$ is the kernel power of the dissipative weight function set to $0.5$ for both unit sets. The complete \textsc{dpd} parameter sets used for \textsc{uca}--water and \textsc{uca}--gas interfaces are given in Table \ref{tab:dpd} of Supplementary Information \ref{sec:dpdinter}.

\subsubsection*{Inducing compression and buckling}
\addcontentsline{toc}{subsubsection}{Inducing compression and buckling}

\noindent
According to the semi-empirical equation of state Eq.\,(\ref{peos}), the fluid pressure is linearly proportional to $a_{ww}$, which is confirmed numerically (see Fig.~\ref{fig:enter-label} in Supplementary Information \ref{sec:pressure}).
Thus, water pressure is controlled by varying the interaction parameter $a_{ww}$ between water beads linearly in time
until it reaches a desired final value, where it remains for the second half of the simulation to allow the structure relax into its final shape. 

Unlike that of water, the compressibility of the gas phase plays an important role, significantly affecting the shell's collapse pressure. 
At the buckling transition, the dominant deformation mode is determined by the bending constant (which is small for thin shells) and is thus sensitive to the compressibility of the interior gas.
At realistic pressure, the compressibility of \textsc{dpd} gas, as follows from the \textsc{dpd} equation of state Eq.\,(\ref{peos}), is anomalously low. To remedy this, we model the gas phase as ideal by setting the interactions between gas beads to 0, which also aligns with the ideal gas behavior in the relevant temperature and pressure range. 
Consequently, the resulting \textsc{dpd} gas pressure is extremely low. To ensure mechanical stability, it is compensated by applying an outward force to each triangular face of the shell, evenly distributed among its three vertices to ensure zero torque on the triangle: $\mathbf{f}_p = -p A \mathbf{n}/3$, where $A$ and $\mathbf{n}$ are the area and inward-pointing normal of the triangle (see Supplementary Information \ref{secA1}), and $p$ is the desired pressure compensation.

\subsection*{Open-boundary molecular dynamics}
\addcontentsline{toc}{subsection}{OBMD}
\label{sec:obmd}

\noindent
To perform non-equilibrium simulations we use open-boundary molecular dynamics (\textsc{obmd}) \cite{sablic2015,DelleSite:2017,Potisk:2023}, which allows imposing momentum and/or heat fluxes at the system’s boundaries. A typical \textsc{obmd} setup consists of three fundamental parts: a central do--the region of interest (\textsc{roi}), and two buffer regions in which particle deletion or insertion is performed. The particle number in the buffer regions is controlled by a feedback algorithm
\begin{equation}
\Delta N = - \frac{\Delta t}{\tau_{B}} (N - \alpha_B N_{0}),
\end{equation}
where $\Delta N$ is the number of particles to be deleted ($\Delta N<0$) or inserted ($\Delta N>0$), 
$\Delta t$ and $\tau_{B}$ are the time step and relaxation time of the buffers, $N$ and $N_0$ the current and equilibrium particle numbers in either buffer, and $\alpha_B$ an empirical customized parameter.
Particle insertion is facilitated by the \textsc{usher} algorithm \cite{Delgado:2003:1}, which employs an iterative steepest descent scheme on the potential energy surface.

\textsc{obmd} imposes boundary conditions by adding external forces $\mathbf{f}_{i}$ to all buffer particles $i$, determined from the momentum balance 
\begin{equation}
    \bm{\mathsf{J}}A \cdot \mathbf{n} = \sum_{i} \mathbf{f}_{i} + \sum_{i'} \frac{\Delta (m_{i'}\mathbf{v}_{i'})}{\Delta t},
    \label{Jrealize}
\end{equation}
where $\bm{\mathsf{J}}$ is the momentum current density tensor, $A$ the surface area of the interface between the buffer and the \textsc{roi}, and $\mathbf{n}$ the normal to this interface pointing towards the center of the \textsc{roi}. The second sum in Eq.\,(\ref{Jrealize}) stands for the momentum gain or loss upon insertion or deletion of particles $i'$ in the current time step. 
The momentum flux across the buffer-\textsc{roi} interface is dictated by the desired boundary condition for the stresses, e.g. on boundaries with normals $\mathbf{n} = \pm \hat{\mathbf{e}}_{x}$:
\begin{equation}
    \bm{\mathsf{J}} \cdot \mathbf{n} = P_{xx}  \mathbf{n} + P_{yx} \hat{\mathbf{e}}_{y} + P_{zx} \hat{\mathbf{e}}_{z},  
\end{equation}
where $P_{xx}$ is the equilibrium pressure, and $P_{yx}$, $P_{zx}$ the shear stress components.

\subsection*{Quasiharmonic analysis}
\addcontentsline{toc}{subsection}{Quasiharmonic analysis}

\noindent
In the analysis of molecular dynamics simulations of biomolecules, extracting eigenmodes and their frequencies is essential. The well-established method for this is quasiharmonic analysis\cite{karplus1981, amadei1993, dusanka1995}, also sometimes referred to as principal component analysis (\textsc{pca}) \cite{Hotelling:1933}. This approach is not limited to molecules but can also be applied to coarse-grained models of elastic objects, such as carbon nanotubes \cite{xu2008, chen2011}, and in our case to \textsc{uca}s.

The thermalized equilibrium configurational probability distribution of membrane vertices is given by
\begin{equation}
    P({\bf x})=\frac{e^{-\beta E(\mathbf{x})}}{\int e^{-\beta E(\mathbf{x})} \mathrm{d}\mathbf{x}},
    \label{eq: gaussian}
\end{equation} 
where $E(\mathbf{x})$ is the potential energy of the configuration expressed by a supervector $\mathbf{x}$ of particle coordinates, and $\beta = 1/(k_BT_0)$.
In the quasiharmonic approximation, the potential energy surface is assumed a quadratic function $E(\mathbf{x}) \approx E(\langle \mathbf{x} \rangle) + \frac12 (\mathbf{x}-\langle\mathbf{x}\rangle)^T \bm{\mathsf{V}}(\mathbf{x}-\langle\mathbf{x}\rangle)$, where $V_{ij} = \frac{\partial^2E}{\partial x_i \partial x_j}$ is the Hessian, and $\langle ~.~ \rangle$ denotes thermal average, which is in our case calculated as a time-average of the vertex positions. 
Within this approximation, the configurational probability function takes the form of a multivariate Gaussian distribution and therefore the Hessian can be extracted from the trajectory through the covariance matrix $\Sigma_{ij} =\left\langle (x_i-\langle x_i \rangle) (x_j-\langle x_j \rangle) \right\rangle$ of particle coordinates:
\begin{equation}
    \bm{\mathsf{V}} = k_BT_0 \bm{\mathsf{\Sigma}}^{-1},
\end{equation}
In determining $\bm{\mathsf{\Sigma}}$, we subtract the motion of the center of mass of the object and remove rotational motion using a trajectory alignment algorithm from the MDAnalysis Python package \cite{Michaud:2011,Gowers:2019}. This algorithm finds the optimal rotation matrix between the current and reference configurations by minimizing the root mean square deviation between the configurations \cite{theobald2005, liu2009}.

Ignoring the damping coming from the solvent and the gas, the equation of motion reads:
\begin{equation}
    \bm{\mathsf{M}}\ddot{\mathbf{x}} + \bm{\mathsf{V}}\mathbf{x} = 0,
\end{equation}
where $\bm{\mathsf{M}}$ is the diagonal mass matrix, $\bm{\mathsf{M}} = m \bm{\mathsf{I}}$ in our case. 
Using the ansatz $\mathbf{x} = \mathbf{x}_0e^{\mathrm{i}\omega t}$, one obtains the generalized eigenvalue problem:

\begin{equation}
    \left(\bm{\mathsf{\Sigma}} - \frac{k_BT_0}{\omega^2}\bm{\mathsf{M}}^{-1}\right)\mathbf{x}_0 = 0,
\end{equation}
which can be recast into a standard eigenvalue problem by defining $\mathbf{q}_0 = \sqrt{\bm{\mathsf{M}}}\mathbf{x}_0$ and $\bm{\mathsf{\Sigma}}' = \sqrt{\bm{\mathsf{M}}}\bm{\mathsf{\Sigma}} \sqrt{\bm{\mathsf{M}}}$:
\begin{equation}
    \left(\bm{\mathsf{\Sigma}}' - \lambda' \bm{\mathsf{I}}\right)\mathbf{q}_0 = 0,
\end{equation}
where the eigenfrequencies of the structural vibrations 
\begin{equation}
    \omega = \sqrt{\frac{k_BT_0}{\lambda'}}
\end{equation}
are obtained from the eigenvalues $\lambda'$.

\backmatter

\bmhead{Code Availability}
The code will be made publicly available in our GitHub repository upon acceptance of the paper.

\bmhead{Acknowledgments}

The authors acknowledge the financial support under ERC Advanced Grant MULTraSonicA (Grant No. 885155) from the European Research Council. The authors also acknowledge the financial support from the Slovenian Research and Innovation Agency (Grant No. P1-0002). Finally, the authors acknowledge the HPC RIVR consortium (www.hpc-rivr.si) and EuroHPC JU (eurohpc-ju.europa.eu) for providing computing resources of the HPC system Vega at the
Institute of Information Science (www.izum.si).

\bmhead{Author contributions}

N.N., M.L. and T.P. performed the simulations. N.N., M.L., D.S., T.P. and M.P. analyzed the results and wrote the paper. N.N., M.L. and T.P. modified the software. M.P. acquired the funding and conceived the study.

\bmhead{Competing interests}
The authors declare no competing interests.

\newpage

\makeatletter
\renewcommand \thesection{S\@arabic\c@section}
\renewcommand\thetable{S\@arabic\c@table}
\renewcommand \thefigure{S\@arabic\c@figure}
\makeatother
\setcounter{figure}{0}

%

\begin{center}
\large{Supplementary Information for:} \\
\Large{Dissipative particle dynamics models of encapsulated microbubbles and gas vesicles for biomedical ultrasound simulations}
\end{center}

\vspace{1cm}
\begin{center}
    {This PDF file includes:}
\end{center}
\begin{itemize}
    \item Supplementary Figures S1-S9 
    \item Supplementary Tables S1-S4 
\end{itemize}
\vspace{0.5cm}

\setcounter{equation}{31}

\section{Elastic energy}\label{appendix:b}

\noindent
Computing the elastic force on a vertex bead of a triangulated object requires an expression for the total elastic energy as a function of the vertex coordinates.
The principle of isotropy of space leads to certain restrictions on the form of the elastic energy of anisotropic objects \cite{boehler1987}. More specifically, a simultaneous rotation of both the body and the deformation $\bm{\varepsilon}$ should lead to the same elastic energy $U$:
\begin{equation} \label{isoen}
    U(\bm{\varepsilon}, \bm{\mathsf{M}}) = U(\bm{\mathsf{Q}}\bm{\varepsilon}\bm{\mathsf{Q}}^T, \bm{\mathsf{Q}}\bm{\mathsf{M}}\bm{\mathsf{Q}}^T) \quad \forall \bm{\mathsf{Q}} \in \mathrm{O}(3). 
\end{equation}
The elastic energy $U$ is therefore a scalar isotropic function of two symmetric second order tensors; the elastic strain $\bm{\varepsilon}$ and the structural tensor $\bm{\mathsf{M}}$. Conversely one can also view the elastic energy as an anisotropic function with respect to the elastic strain $\bm{\varepsilon}$:
\begin{equation}
    U(\bm{\varepsilon}, \bm{\mathsf{M}}) = U(\bm{\mathsf{Q}}\bm{\varepsilon}\bm{\mathsf{Q}}^T, \bm{\mathsf{M}}) \quad\forall \bm{\mathsf{Q}} \in \mathcal{G} \subseteq \mathrm{O}(3), \label{invariance}
\end{equation}
where $\mathcal{G}$ denotes the material symmetry group. The main task is then to find all the possible scalar terms that are invariant under the action of all the group elements $\bm{\mathsf{Q}}\in\mathcal{G}$.

The elastic energy $U$ can also be derived by first writing down the elastic tensor $C_{ijkl}^{3D}$, which relates the strain tensor $\varepsilon_{ij}^{3D}$ to the elastic stress tensor $\sigma_{ij}^{3D}$,
\begin{equation}\label{strstr}
    \sigma_{ij}^{3D} = C_{ijkl}^{3D}\varepsilon_{kl}^{3D}.
\end{equation}
In linear elastic theory of thin shells, in-plane and bending deformations are decoupled owing to the assumption that normal stresses are negligible compared to internal tangential stresses (even if bending is induced by normal stress). Under this approximation, the 3D deformation field of the thin shell can be split into in-plane ($xy$) strain $\varepsilon_{ij}$ and bending $H_{ij}$,
\begin{equation}\label{streq}
    \varepsilon_{ij}^{3D} = \varepsilon_{ij} - zH_{ij}.
\end{equation}
The total elastic energy is then simply a sum of the in-plane elastic ($U_{el}$) and bending ($U_b$) contributions:
\begin{align}
 U &= \int_V\! {\textstyle\frac12} C_{ijkl}^{3D}\varepsilon_{ij}^{3D}\varepsilon_{kl}^{3D}\,\mathrm{d}V \nonumber \\
  &= \int\!\!\! \int_{z=-h/2}^{h/2}{\textstyle\frac12} C_{ijkl}^{3D}\varepsilon_{ij}\varepsilon_{kl}\,\mathrm{d}A\,\mathrm{d}z + \int\!\!\! \int_{z=-h/2}^{h/2}{\textstyle\frac12} z^2C_{ijkl}^{3D}H_{ij}H_{kl}\,\mathrm{d}A\,\mathrm{d}z \nonumber \\
    &= \underbrace{\int\! {\textstyle\frac12} C_{ijkl}\varepsilon_{ij}\varepsilon_{kl}\,\mathrm{d}A}_{U_{el}} + \underbrace{\int\! {\textstyle\frac12} D_{ijkl}H_{ij}H_{kl}\,\mathrm{d}A}_{U_b},
\end{align}
where we have identified the two-dimensional in-plane elastic tensor $C_{ijkl}$ and the flexural rigidity tensor $D_{ijkl}$ as integrals of the full 3D elastic tensor $C_{ijkl}^{3D}$ over the thickness $h$ of the shell,
\begin{eqnarray} 
    C_{ijkl} &=& \int_{z=-h/2}^{h/2}\! C_{ijkl}^{3D}\,\mathrm{d}z = C_{ijkl}^{3D}\,h \label{cijkla} \\
    D_{ijkl} &=& \int_{z=-h/2}^{h/2}\! z^2C_{ijkl}^{3D}\,\mathrm{d}z = C_{ijkl}^{3D}\,\frac{h^3}{12}. \label{dijkla}
\end{eqnarray}
It should be emphasized that, in general, thin shell theory is not always applicable, particularly for single-molecule layered membranes \cite{evans1980}, and the bending coefficients should, in principle, be regarded as independent of the in-plane elastic ones. 

\paragraph{In-plane elastic energy}
In the constant strain triangle approximation (CST) \cite{Turner:1956}, where the strain field $\bm{\varepsilon}$ is constant across each of the triangles, in-plane elastic energy $U_{el}$ can be written as a sum over triangles of the triangulated surface, 

\begin{equation}
    U_{el} = \int{\textstyle\frac12} C_{ijkl}\varepsilon_{ij}\varepsilon_{kl}\,\mathrm{d}A \to \sum_t {\textstyle\frac12} C_{ijkl}^t \varepsilon_{ij}^t  \varepsilon_{kl}^t \,A_0^t, \label{inplane}
\end{equation}
where the sum runs through all the triangles of the object and the superscript $t$ denotes the corresponding quantity evaluated on the $t$-th triangle. Since the strains are calculated relative to the reference stress-free configuration, the $A_0^t$ denotes the stress-free area of the $t$-th triangle. 

For an isotropic material, the elastic tensor $C_{ijkl}$ is expressed in terms of the Kronecker delta $\delta_{ij}$ Eq.\,(\ref{isotensor}) in the main text, from which the in-plane elastic energy $U_{el}^{iso}$ is then derived using Eq.\,(\ref{Uel-general}).
However, since isotropic elasticity has been widely used in literature \cite{evans1980,economides2021}, we instead adopt one of the established expressions for the isotropic elastic energy \cite{Lim:2008}, which is also included in the Mirheo simulation package~\cite{alexeev2020}:
\begin{equation} \label{iso-elast}
    U_{el}^{iso} = \frac{K_a}{2} \sum_{t=1}^{N_t} A_t^0[\alpha_t^2 + a_3 \alpha_t^3 + a_4 \alpha_t^4] + \mu \sum_{t=1}^{N_t} A_t^0[\beta_t + b_1 \alpha_t \beta_t + b_2 \beta_t^2], \end{equation}
where $\alpha_t$ and $\beta_t$ are the elastic invariants corresponding to the triangular face $t$ (see Fig~\ref{fig-triangle}). This expression also includes nonlinear terms (coefficients $a_3, a_4, b_1$, $b_2$). 
The invariants $\alpha$ and $\beta$ are connected with compression and shear deformation, respectively:
\begin{equation}
 \alpha = \lambda_1\lambda_2 - 1, \qquad   
 \beta = \frac{\lambda_1^2+\lambda_2^2}{2\lambda_1\lambda_2}-1,  
\end{equation}
with $\lambda_{1,2}$ the eigenvalues of the deformation gradient $F_{ij}$, corresponding to local principal stretches \cite{evans1980,Lim:2008} of a deformed triangular element. 
Details on the calculation of the deformation gradient tensor are given in Sec.~(\ref{secA1}). The nonlinear coefficients $a_3, a_4, b_1$ and $b_2$ are often used when describing the elastic properties of \textsc{rbc}s, especially under large strains. In this work, all nonlinear coefficients are set to zero.

For small strains, Eqs.\,(\ref{iso-elast}) and (\ref{inplane}) are equivalent, as we show next.
We express the eigenvalues $\lambda_{1,2}$ in terms of eigenvalues $e_{1,2}$ of the strain tensor $\bm{\varepsilon}$, i.e., the principal strains, and perform a Taylor expansion around zero strain:
\begin{align}
\alpha &= \lambda_1\lambda_2-1 = \sqrt{1+2e_1}\sqrt{1+2e_2}-1 \nonumber \\
&= \sqrt{1+2\mathrm{tr}(\bm{\varepsilon})+2(\mathrm{tr}(\bm{\varepsilon})^2-\mathrm{tr}(\bm{\varepsilon}^2))}-1  \nonumber \\&\approx \mathrm{tr}(\bm{\varepsilon}) + O(|\bm{\varepsilon}|^2), \\
\beta &= \frac{\lambda_1^2+\lambda_2^2}{2\lambda_1\lambda_2}-1 = \frac{1+2e_1+1+2e_2}{2\sqrt{1+2(e_1+e_2)+e_1e_2}}-1 \nonumber \\
& = \frac{2+2\mathrm{tr}(\bm{\varepsilon})}{2 \sqrt{1+2\mathrm{tr}(\bm{\varepsilon})+2(\mathrm{tr}(\bm{\varepsilon})^2-\mathrm{tr}(\bm{\varepsilon}^2))}}-1\nonumber \\
& \approx \mathrm{tr}(\bm{\varepsilon}^2)-\frac{1}{2}\mathrm{tr}(\bm{\varepsilon})^2+ O(|\bm{\varepsilon}|^3).
\end{align}
The quantities $\mathrm{tr}(\bm{\varepsilon})$ and $\mathrm{tr}(\bm{\varepsilon}^2)-\frac{1}{2}\mathrm{tr}(\bm{\varepsilon})^2$ are exactly the strain tensor invariants corresponding to area dilatation and pure shear deformations, respectively.

For an orthotropic material, the elastic tensor $C_{ijkl}$ is constructed using a structural tensor $M_{ij} = m_im_j$, which describes the preferred direction $\mathbf{m}$ (normalized vector), along with the Kronecker delta $\delta_{ij}$ (or transverse Kronecker delta $\delta_{ij}^{\perp} = \delta_{ij}-m_im_j$). 
The tensor $C_{ijkl}$ should be invariant with respect to the inversion $\mathbf{m}\to -\mathbf{m}$. Taking into account these symmetries one gets four independent terms (Eq.\,(\ref{tensor}) in the main text),
\begin{align} \label{tensor}
C_{ijkl} &= K_a \delta_{ij}\delta_{kl} + \mu\left(\delta_{ik}\delta_{jl}+\delta_{il}\delta_{jk}-\delta_{ij}\delta_{kl}\right) \nonumber\\ &+ (\mu_L-\mu)(m_im_l \delta_{jk} + m_j m_l \delta_{ik} + m_i m_k \delta_{jl} + m_jm_k \delta_{il}) \nonumber\\
&+ c m_im_jm_km_l,
\end{align}
with $\mu_L$, $c$ the anisotropic elastic coefficients. 
Since the \textsc{gv} rotates and deforms over time, the axis $\mathbf{m}$ is updated at each step, simply by tracking the end points of the \textsc{gv}. See Section \ref{secA1} for details on the calculation of $\mathbf{m}$.

Thus, the complete expression for the in-plane elastic energy of the orthotropic material is
\begin{align} \label{gvelastan}
U_{el}^{aniso} = U_{el}^{iso} + \sum_{t=1}^{N_t} A_0^t\left[2(\mu_L-\mu)I_3 + \frac12 c I_4^2\right],
\end{align}
where we introduced scalar strain invariants for compactness:
\begin{align} 
    I_3 &= \mathbf{m}^T\bm{\varepsilon}^T \bm{\varepsilon} \mathbf{m}, \label{invariants1} \\ 
    I_4 &= \mathbf{m}^T\bm{\varepsilon} \mathbf{m}.  \label{invariants2}
\end{align}
For vanishing anisotropy, where $\mu_L - \mu = 0$, $c = 0$, isotropic elasticity is recovered in Eq.\,(\ref{gvelastan}).
In principle, there is one more term quadratic in $\bm{\varepsilon}$, proportional to $I_5 \equiv \alpha I_4 \approx \mathrm{tr}(\bm{\varepsilon})I_4$ or equivalently, a term proportional to $(\delta_{ij}m_km_l+\delta_{kl}m_i m_j)$ in Eq.\,(\ref{tensor}). In 2D, it can be expressed as a sum of other invariants in the limit of low strain and was therefore omitted.

In practice, it is sometimes more convenient to work with the so-called engineering constants, which are typically easier to measure and are therefore more frequently reported in experimental literature. They are defined in the small-strain limit through the relations between the strain $\varepsilon_{ij}$ and stress $\sigma_{ij}$ components. For 2D isotropic elasticity the definition is
\begin{equation} \label{eqmat-iso}
\begin{bmatrix} \varepsilon_{11} \\ \varepsilon_{22} \\ \varepsilon_{12} \end{bmatrix} =
\begin{bmatrix} \frac{1}{E} & -\frac{\nu}{E} & 0 \\  -\frac{\nu}{E} & \frac{1}{E} & 0 \\
0 & 0 & \frac{1}{2G} \end{bmatrix}
\begin{bmatrix} \sigma_{11} \\ \sigma_{22} \\ \sigma_{12} \end{bmatrix},
\end{equation}
where $E$, $\nu$, and $G=\frac{E}{2(1+\nu)}$ are the Young's modulus, Poisson's ratio, and shear modulus, respectively. 
The bulk $K_a$ and shear $\mu$ moduli in Eq.\,(\ref{isotensor}) are expressed as
\begin{align}
K_a &= \frac{E}{2(1-\nu)}\\
\mu &= G = \frac{E}{2(1+\nu)},
\end{align}
while the converse relations are:
\begin{align}
    E &= \frac{4K_a\mu}{K_a + \mu} \label{Eeng}\\
    \nu &= \frac{K_a - \mu}{K_a + \mu} \label{nueng}.
\end{align}

In 2D orthotropic elasticity the relation between $\varepsilon_{ij}$ and $\sigma_{ij}$ involves two additional engineering constants:
\begin{equation} \label{eqmat}
    \begin{bmatrix} \varepsilon_{11} \\ \varepsilon_{22} \\ \varepsilon_{12} \end{bmatrix} =
    \begin{bmatrix} \frac{1}{E_1} & -\frac{\nu_{lt}}{E_1} & 0 \\  -\frac{\nu_{lt}}{E_1} & \frac{1}{E_2} & 0 \\
    0 & 0 & \frac{1}{2G} \end{bmatrix}
    \begin{bmatrix} \sigma_{11} \\ \sigma_{22} \\ \sigma_{12} \end{bmatrix},
\end{equation}
where $\nu_{lt}=-\frac{\varepsilon_{22}}{\varepsilon_{11}}$ is the Poisson's ratio for uniaxial stress $\sigma_{11}$. With that, the other Poisson's ratio $\nu_{tl}=-\frac{\varepsilon_{11}}{\varepsilon_{22}}$ for uniaxial stress $\sigma_{22}$ is also set,
\begin{equation}
    \frac{\nu_{lt}}{E_l} = \frac{\nu_{tl}}{E_t}, \label{poissonlt}
\end{equation}
which was already taken into account in Eq.\,(\ref{eqmat}).

The four engineering constants
can be used to calculate the elastic coefficients in Eq.\,(\ref{tensor}):
\begin{align} 
K_a &= \frac{E_tE_l(1+\nu_{lt})}{2(E_l-\nu_{lt}^2E_t)} \label{Kmu-eng}\\
\mu &= \frac{E_tE_l(1-\nu_{lt})}{2(E_l-\nu_{lt}^2E_t)}\\
\mu_L &= G\\
c &= \frac{E_l^2+E_lE_t+4E_tG\nu_{lt}^2-4E_lG-2E_lE_t\nu_{lt}}{E_l-\nu_{lt}^2E_t}.
\end{align}
The converse relations are
\begin{align} 
E_t &= \frac{4 \Lambda \mu_{L} + \Lambda c + 8 \mu_{L} \mu - 4 \mu^{2} + 2 \mu c}{\Lambda + 4 \mu_{L} - 2 \mu + c} \\
E_l &=\frac{4 \Lambda \mu_{L} + \Lambda c + 8 \mu_{L} \mu - 4 \mu^{2} + 2 \mu c}{\Lambda + 2 \mu}\\
\nu_{lt} &= \frac{\Lambda}{\Lambda + 2 \mu_{L}} \\
G &= \mu_L, \label{eng-Kmu}
\end{align}
where $\Lambda\equiv K_a-\mu$ is the Lame's first parameter. Thermodynamic stability of the 2D orthotropic material requires $E_t, E_l, G > 0$ and $|\nu_{lt}| < \sqrt{\frac{E_l}{E_t}}$.

\paragraph{Bending energy}

\noindent
The bending energy Eq.\,(\ref{gv-bend-ene}) in the main text is perhaps easier to understand by comparing it to the bending of a flat plate with a particular direction of the anisotropy axis (longitudinal direction), e.g., ${\mathbf{m}} = \hat{\mathbf{e}}_x$: 
\begin{equation} \label{gv-bend}
    U_b({\mathbf{m}} = \hat{\mathbf{e}}_x, B_{ij} = 0) = \frac12 D_{xx}H_{xx}^2 + D_{xy}H_{xx}H_{yy} + \frac12 D_{yy}H_{yy} + \frac12 D_G H_{xy}^2,
\end{equation}
where $D_{xx} = \frac{E_lh^2}{12(1-\nu_{tl}\nu_{lt})}$, $D_{yy} = \frac{E_th^2}{12(1-\nu_{tl}\nu_{lt})}$, $D_{xy} = \frac{E_t\nu_{lt}h^2}{12(1-\nu_{tl}\nu_{lt})}$ and $D_{G} = \frac{Gh^2}{12}$ are the flexural rigidities. Clearly, bending along the anisotropy axis $x$ depends on $E_l$, while bending along $y$ depends on $E_t$. 
Given that \textsc{gv}s are stiffer along their axis, one expects circumferential bending to be more easily induced than axial bending. 
This is also confirmed in our numerical calculations of compression and buckling, as well as in our eigenmode analysis.

There are multiple ways of discretizing the bending energy \cite{Bian2020, gekle2017}, such as the Monge-form based discretization by Kantor and Nelson \cite{kantor1987}, the edge curvature approach by Jülicher \cite{juelicher1996}, and approaches that discretize the Laplace-Beltrami operator \cite{gompper1996, meyer2003}. We use the Kantor-Nelson form, which takes into account the angle between each pair of triangles:
\begin{equation} \label{Ub-t}
    U_b = \kappa_b\sum_{\langle i,j\rangle} \left[1-\cos(\theta_{ij}-\theta_{ij}^0)\right],
\end{equation}
where $\kappa_b$ is the bending parameter, while $\theta_{ij}$ and $\theta_{ij}^0$ are the actual and spontaneous angles between two adjacent triangles, respectively. To reproduce Eq.\,(\ref{iso-bend}), $\kappa_b = \frac{2}{\sqrt{3}}\kappa$ must be used. For simplicity, we use Eq.\,(\ref{Ub-t}) for the bending energy of both \textsc{emb}s and \textsc{gv}s. Anisotropic bending effects are retained due to the anisotropic in-plane elasticity generally associated with the bending of shells.

\section{Elastic forces between mesh vertices}\label{secA1}

\noindent
Elastic forces acting on the vertex nodes are calculated as derivatives of the total elastic energy $U = U_{el} + U_b$ with respect to the vertex coordinates. In Mirheo \cite{alexeev2020}, each GPU thread is mapped to one vertex and loops over all adjacent triangles of that vertex. Therefore, only the force on one of the vertices of a given triangle needs to be specified, e.g.,
\begin{align} \label{eqaniso}
    \mathbf{f}_1 =& -\frac{\partial U}{\partial \mathbf{x}_1}.
\end{align}

\begin{figure}[!htbp]
    \centering
    \includegraphics[width=0.95\linewidth]{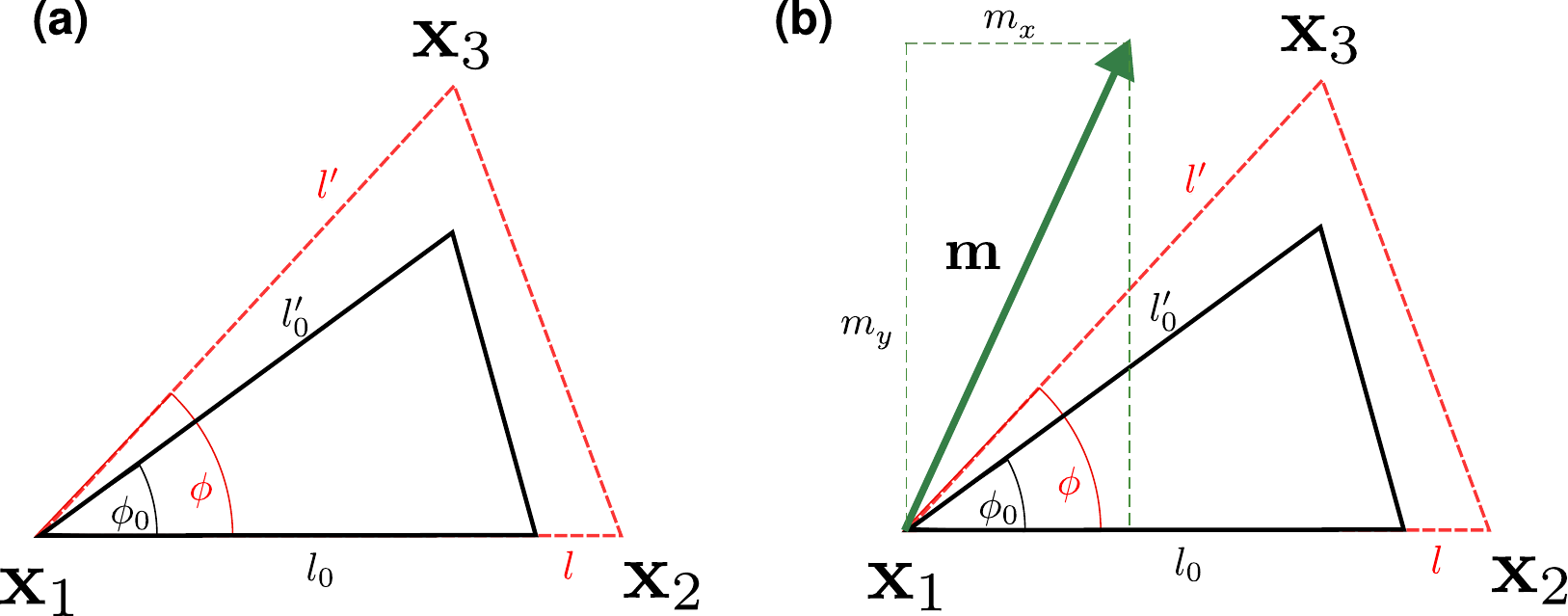}
    \caption{a) A deformed triangular face of an object (red dashed lines) superimposed to a reference triangular face (black solid lines) spanned by vertices $\mathbf{x}_1$, $\mathbf{x}_2$ and $\mathbf{x}_3$. b) A scheme representing the projected anisotropy axis $\mathbf{m}$ onto a given triangular element together with its two components $m_x$ and $m_y$.}
    \label{fig-triangle}
\end{figure}

An analytical expression for the deformation gradient $\bm{\mathsf{F}}$ entering $U$ is thus required.
The easiest way to calculate the deformation gradient for a \textsc{cst} triangular element is to first super-impose one of the sides of a triangle (e.g., $\boldsymbol{\xi}_{12} = \mathbf{x}_2 - \mathbf{x}_1$) and express the coordinates of the deformed configurations ($\mathbf{x}_2$ and $\mathbf{x}_3$) in terms of their reference counterparts ($\mathbf{x}_2^0$ and $\mathbf{x}_3^0$), see Fig.~\ref{fig-triangle}. Here the basis vectors are $\hat{\mathbf{e}}_1 = \frac{\bxi_{12}}{|\bxi_{12}|}$ and $\hat{\mathbf{e}}_2 = \frac{\mathbf{n}\times \bxi_{12}}{|\mathbf{n}\times \bxi_{12}|}$. The deformation gradient then reads \cite{Lim:2008}
\begin{equation} \label{defgrad}
\bm{\mathsf{F}} = \begin{bmatrix}
  a & b \\
  0 & c
\end{bmatrix},
\end{equation}
where $a$, $b$, and $c$ are determined by comparing the deformed and reference triangular faces, see Fig.~\ref{fig-triangle}:
\begin{align}
a &= \frac{l}{l_0} = \frac{\sqrt{\bxi_{12}\cdot\bxi_{12}}}{l_0}, \\
b &= \frac{1}{\sin(\phi_0)}\left(\frac{l'}{l_0'}\cos(\phi)-\frac{l}{l_0}\cos(\phi_0)\right) =  \frac{1}{\sin(\phi_0)}\left(\frac{\bxi_{12}\cdot\bxi_{13}}{ll_0'}-a\cos(\phi_0)\right) \nonumber\\
  &= \frac{l_0l_0'}{2A_0}\left(\frac{\bxi_{12}\cdot\bxi_{13}}{ll_0'}-a\frac{\bxi_{12}^0\cdot \bxi_{13}^0}{l_0l_0'}\right)=\frac{1}{2A_0}\left(\frac{\bxi_{12}\cdot\bxi_{13}}{a}-a\bxi_{12}^0\cdot \bxi_{13}^0\right), \\ 
c &= \frac{l'}{l_0'}\frac{\sin(\phi)}{\sin(\phi_0)}= \frac{\sqrt{(\bxi_{12}\times\bxi_{13})\cdot(\bxi_{12}\times\bxi_{13})}}{ll_0'\sin(\phi_0)} = \frac{2A}{ll_0'\sin(\phi_0)}=\frac{A}{aA_0},
\end{align}
where $A$ denotes the deformed area of the triangle: $A = \frac{1}{2}|\mathbf{n}^*|$, 
with $\mathbf{n}^* = \bxi_{12}\times\bxi_{13}$, $\phi_0, \phi$ are the angles between the edges $\boldsymbol{\xi}_{12}$ and $\boldsymbol{\xi}_{13}$, $l_0, l$ are the lengths of the edge $\boldsymbol{\xi}_{12}$, while $l_0', l'$ are the lengths of the edge $\boldsymbol{\xi}_{13}$. The quantities calculated in the reference configuration have a subscript $0$. 
We note that the triangles are indexed so that the vectors $\mathbf{n}^*$ point inside the object.

\noindent
Due to the involved expressions for the total force $\mathbf{f}_1$ on a given vertex, we split this force into two contributions; one from the in-plane elastic energy, $\mathbf{f}_1^{el}=-\frac{\partial U^{el}}{\partial \mathbf{x}_1}$ and the other from the bending energy, $\mathbf{f}_1^{b}=-\frac{\partial U^{b}}{\partial \mathbf{x}_1}$. For the orthotropic case, the force $\mathbf{f}_1^{el}$ reads

\begin{align} \label{eqaniso}
\mathbf{f}_1^{el} =& -\frac{\partial U_{el}}{\partial \mathbf{x}_1} = \mathbf{f}_1^{el,iso} - \sum_{t=1}^{N_t} A_0^t\left[2(\mu_L-\mu) \frac{\partial I_3}{\partial \mathbf{x}_1} + cI_4 \frac{\partial I_4}{\partial \mathbf{x}_1}\right],
\end{align}
where $I_3$, $I_4$ are the scalar invariants Eqs.\,(\ref{invariants1})-(\ref{invariants2}), and $\mathbf{f}_1^{el,iso}$ denotes the force when only the isotropic part of the energy Eq.\,(\ref{iso-elast}) is included. We have implemented the orthotropic elastic force as an additional membrane force kernel in Mirheo.

The relevant derivatives with respect to the the vertex coordinates, used in numerical computation of Eq.\,(\ref{eqaniso}), are:
\begin{align} 
\frac{\partial A}{\partial \mathbf{x}_1} &= - \frac{1}{4A}\mathbf{n}^*\times \bxi_{32}, \\ 
\frac{\partial a}{\partial \mathbf{x}_1} &= -\frac{\bxi_{12}}{ll_0}, \\
\frac{\partial a}{\partial \mathbf{x}_2} &= \frac{\bxi_{12}}{ll_0}, \\
\frac{\partial a}{\partial \mathbf{x}_3} &= 0, \\
\frac{\partial b}{\partial \mathbf{x}_1} &= \frac{1}{\sin\phi_0}\left(-\frac{\bxi_{13}}{ll_0'}-\frac{\bxi_{12}}{ll_0'}+\frac{\bxi_{12}\cdot \bxi_{13}}{l^3l_0'}\bxi_{12}+\frac{\cos\phi_0}{ll_0}\bxi_{12}\right), \\
\frac{\partial b}{\partial \mathbf{x}_2} &= \frac{1}{\sin\phi_0}\left(\frac{\bxi_{13}}{ll_0'}-\frac{\bxi_{12}\cdot \bxi_{13}}{l^3l_0'}\bxi_{13}-\frac{\cos\phi_0}{ll_0}\bxi_{13}\right), \\
\frac{\partial b}{\partial \mathbf{x}_3} &= \frac{1}{\sin\phi_0}\frac{\bxi_{12}}{ll_0'}, \\
\frac{\partial c}{\partial \mathbf{x}_1} &= \frac{\mathbf{n}^*\times \bxi_{23}}{2A ll_0'\sin\phi_0}+\frac{2A}{l^3l_0'\sin\phi_0}\bxi_{12}, \\
\frac{\partial c}{\partial \mathbf{x}_2} &= \frac{\mathbf{n}^*\times \bxi_{13}}{2A ll_0'\sin\phi_0}-\frac{2A}{l^3l_0'\sin\phi_0}\bxi_{12}, \\
\frac{\partial c}{\partial \mathbf{x}_3} &= \frac{\mathbf{n}^*\times \bxi_{12}}{2A ll_0'\sin\phi_0}.
\end{align}

The invariants in Eq.\,(\ref{eqaniso}) and their derivatives with respect to the vertex coordinate can be more compactly expressed using the invariants involving the the axis of anisotropy $\mathbf{m}$ and the right Cauchy strain tensor $\bm{\mathsf{C}}$
\begin{align}
    I_3^C &= \mathbf{m}^T\bm{\mathsf{C}}^T\bm{\mathsf{C}}\mathbf{m} = 4I_3+4I_4+1 =  m_x^2(a^4+a^2b^2)+2m_xm_y(a^3b+ab^3+abc^2)\nonumber \\
    &+ m_y^2(a^2b^2+(b^2+c^2)^2),\\
    I_4^C &= \mathbf{m}^T\bm{\mathsf{C}}\mathbf{m} = 2I_4+1= m_x^2a^2+2m_xm_y ab + m_y^2(b^2+c^2),
\end{align}
where
\begin{equation} \label{cauchy}
    \bm{\mathsf{C}} = \bm{\mathsf{F}}^T\bm{\mathsf{F}} = \begin{bmatrix}
  a^2 & ab \\
  ab & b^2+c^2
\end{bmatrix}.
\end{equation}
The derivatives of the invariants $I_3^C$ and $I_4^C$ are:
\begin{align}
\frac{\partial I_3^C}{\partial \mathbf{x}_1} &= m_x^2\left(4a^3\frac{\partial a}{\partial \mathbf{x}_1}+2ab^2\frac{\partial a}{\partial \mathbf{x}_1}+2a^2b\frac{\partial b}{\partial \mathbf{x}_1}\right) + 2m_xm_y\frac{\partial a}{\partial \mathbf{x}_1}\left(3a^2b+b^3+bc^2\right) \nonumber \\
&+ 2m_xm_y\frac{\partial b}{\partial \mathbf{x}_1}\left(a^3+3ab^2+ac^2\right)+4m_xm_y\frac{\partial c}{\partial \mathbf{x}_1}abc \nonumber \\
&+ 2m_y^2\frac{\partial a}{\partial \mathbf{x}_1}ab^2 + m_y^2\frac{\partial b}{\partial \mathbf{x}_1}(2a^2b+4(b^2+c^2)b) + 4m_y^2\frac{\partial c}{\partial \mathbf{x}_1}(b^2+c^2)c, \\
\frac{\partial I_4^C}{\partial \mathbf{x}_1} &= 2m_x^2a \frac{\partial a}{\partial \mathbf{x}_1} + 2m_xm_y b \frac{\partial a}{\partial \mathbf{x}_1}+ 2m_xm_y a \frac{\partial b}{\partial \mathbf{x}_1}+ 2m_y^2 \frac{\partial b}{\partial \mathbf{x}_1}b+ 2m_y^2 \frac{\partial c}{\partial \mathbf{x}_1}c.
\end{align}

\begin{figure}[!htbp]
    \centering
    \includegraphics[width=0.5\linewidth]{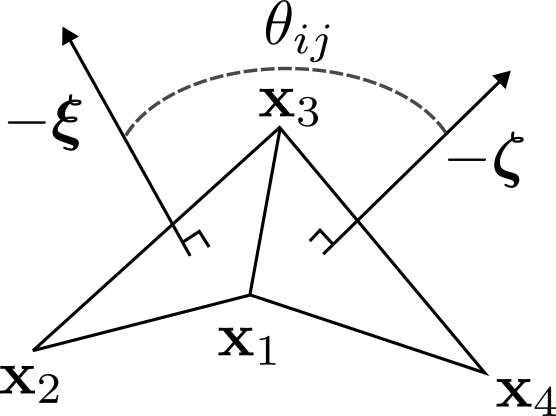}
    \caption{Vertices, normals $\boldsymbol{\xi}$ and $\boldsymbol{\zeta}$, and the angle $\theta_{ij}$ between the adjacent triangles for the calculation of bending forces. For clarity, the inverse normals are shown. Note the interchange of $\mathbf{x}_1$ and $\mathbf{x}_2$ between our indexing convention and that of Ref~\cite{Fedosov:2010thesis}. In our convention, the normals point inward.} 
    \label{fig:dihedral}
\end{figure}

The vector $\mathbf{m}$ is defined locally for each triangle and is written in the same basis as the deformation gradient Eq.\,(\ref{defgrad}).
This vector is derived from the main axis of the gas vesicle $\mathbf{m}^*$, which is for simplicity calculated from the positions of the tips of the conical ends $\mathbf{v}_t$ and $\mathbf{v}_b$, 

\begin{align}
\mathbf{m}^* &= \mathbf{v}_t - \mathbf{v}_b,
\end{align}
where the two choices of assigning $\mathbf{v}_t$ and $\mathbf{v}_b$ are equivalent due to the $\mathbf{m}^* \to -\mathbf{m}^*$ symmetry. 
We project $\mathbf{m}^*$ onto the triangle by subtracting the component along the triangle unit normal $\mathbf{n}$:
\begin{align}
\mathbf{m}^\parallel &= \mathbf{m}^* - \mathbf{n}(\mathbf{n}\cdot \mathbf{m}^*), \\ 
\mathbf{n} &= \frac{\bxi_{12}\times \bxi_{13}}{|\bxi_{12}\times \bxi_{13}|}.
\end{align}
The components of the projected vector $\mathbf{m}^\parallel$ in the basis of the deformation gradient Eq.~(\ref{defgrad}), then read
\begin{align} 
m_x &= \frac{\mathbf{m}^\parallel \cdot \bxi_{12}}{|\mathbf{m}^\parallel||\bxi_{12}|}, \\ 
m_y &= \frac{\mathbf{m}^\parallel \cdot (\mathbf{n}\times \bxi_{12})}{|\mathbf{m}^\parallel||\mathbf{n}\times \bxi_{12}|}.
\end{align}
Calculating the bending forces on the vertices requires iterating through all the pairs of triangles of the triangulated mesh, see Fig.~\ref{fig:dihedral}. This is a consequence of the Kantor-Nelson discretization of the bending energy Eq.\,(\ref{Ub-t}), which takes into account the angles $\theta_{ij}$ between the pairs of adjacent triangles $i$ and $j$. The nodal bending forces on all the vertices $\mathbf{x}_1, \mathbf{x}_2, \mathbf{x}_3, \mathbf{x}_4$ of two adjacent triangles read \cite{Fedosov:2010thesis}
\begin{align}
\mathbf{f}_1^b &=-\frac{\partial U_b}{\partial \mathbf{x}_1}= b_{11}\bxi \times \bxi_{32}+b_{12}\left(\bxi\times \bxi_{43}+\bzeta \times \bxi_{32}\right)+b_{22}\bzeta \times \bxi_{43}, \\
\mathbf{f}_2^b &=-\frac{\partial U_b}{\partial \mathbf{x}_2}= b_{11}\bxi \times \bxi_{13} + b_{12}\bzeta \times \bxi_{13}, \\
\mathbf{f}_3^b &=-\frac{\partial U_b}{\partial \mathbf{x}_3}=b_{11}\bxi \times \bxi_{21}+b_{12}\left(\bxi\times \bxi_{14}+\bzeta \times \bxi_{21}\right)+b_{22}\bzeta \times \bxi_{14}, \\
\mathbf{f}_4^b &=-\frac{\partial U_b}{\partial \mathbf{x}_4}=b_{11}\bxi \times \bxi_{31} + b_{12}\bzeta \times \bxi_{31},
\end{align}
where $b_{12}= -\beta_b \cos(\theta_{ij})/|\bxi|^2$, $b_{12}=\beta_b/(|\bxi||\bzeta|)$, $b_{22}= -\beta_b \cos(\theta_{ij})/|\bzeta|^2$ and $\beta_b = \kappa_b\left(\sin \theta_{ij} \cos\theta_{ij}^0 - \cos \theta_{ij} \sin\theta_{ij}^0\right)$. The normals of the triangles are $\bxi = \bxi_{12} \times \bxi_{13}$ and $\bzeta = \bxi_{41} \times \bxi_{43}$, see Fig.~\ref{fig:dihedral}. The cosine of the angle between the adjacent triangles is determined from the scalar products of the normals, $\cos \theta_{ij} = \frac{\bxi \cdot \bzeta}{|\bxi||\bzeta|}$, which leads to $\sin \theta_{ij} = \pm \sqrt{1-\cos^2 \theta_{ij}}$, where the positive sign is taken if $(\bxi-\bzeta)\cdot \bxi_{24}>0$ and negative otherwise.

\section{Mechanical properties - technical details}
\subsection{Numerical validation of elastic properties}

To validate that the elastic membranes of the ultrasound contrast agents reproduce the desired elastic properties, such as Young's moduli and Poisson's ratios, we test the behavior of a rectangular sheet under extensional stresses $\sigma_{xx}$ and $\sigma_{yy}$ and measure their strains $\varepsilon_{xx}$ and $\varepsilon_{yy}$. To measure the shear modulus $G$, we impose a shear strain and measure the resulting shear stress response.

In these validation simulations, the axis of anisotropy is $\mathbf{m} = \hat{\mathbf{e}}_x$, and the engineering constants are set to $E_t = 50\,k_BT_0/r_c^2$, $E_l = 250\,k_BT_0/r_c^2$, $\nu_{lt} = 0.8$, $G = 5.7\,k_BT_0/r_c^2$.

\begin{figure}[!htbp]
\centering
\begin{minipage}[t]{0.3\textwidth}
\centering
    \includegraphics[width=\textwidth]{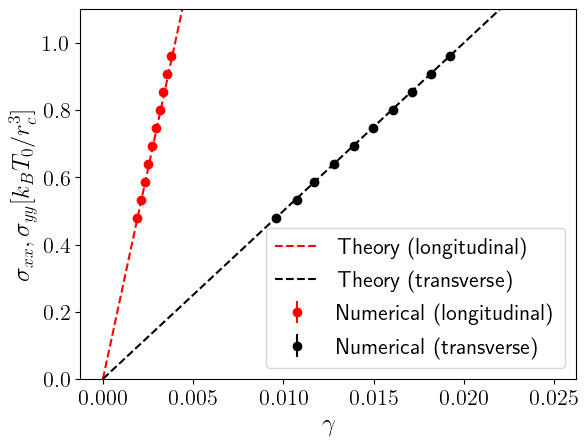}
\end{minipage}
\begin{minipage}[t]{0.3\textwidth}
\centering
    \includegraphics[width=\textwidth]{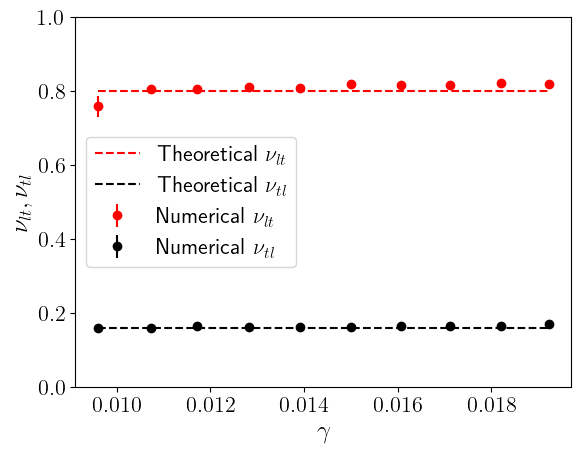}
\end{minipage}
\begin{minipage}[t]{0.3\textwidth}
\centering
    \includegraphics[width=\textwidth]{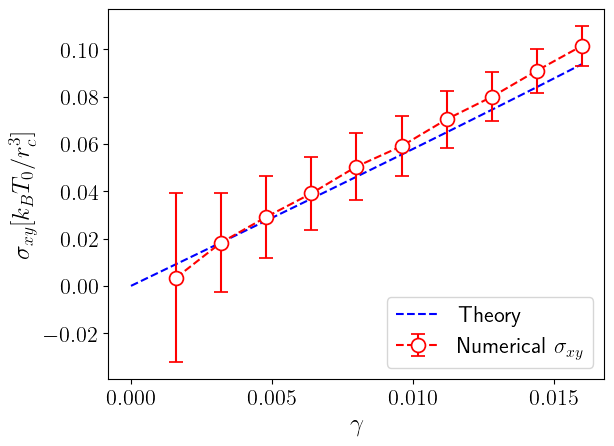}
\end{minipage}
\caption{(a) Dilatational stresses $\sigma_{xx}$ and $\sigma_{yy}$ as functions of strain when stretching the sheet along the $x$ and $y$ axis. (b) Poisson's ratios $\nu_{lt}$ and $\nu_{tl}$ as a function of strain. (c) Shear stress $\sigma_{xy}$ as a function of shear strain. Dashed lines represent theoretical values, set as simulation parameters, and points represent their measured values.}
\label{fig: yng}
\end{figure}

As seen in Fig.~\ref{fig: yng}, the measured Young's moduli, Poisson's ratios $\nu_{lt}$ and $\nu_{tl}$, and shear modulus $\mu_L$ agree excellently with the simulation input values. To calculate the Poisson's ratio, we measure $\nu_{lt} = -\frac{\varepsilon_{yy}}{\varepsilon_{xx}}$ for stretching forces along the $x$ axis, as well as its reciprocal equivalent $\nu_{tl} = -\frac{\varepsilon_{xx}}{\varepsilon_{yy}}$ for stretching along the $y$ axis. The two Poisson's ratios are related according to Eq.\,(\ref{poissonlt}), yielding $\nu_{tl} = 0.16$.

\begin{figure}[!htbp]
    \centering
    \includegraphics[width=0.95\linewidth]{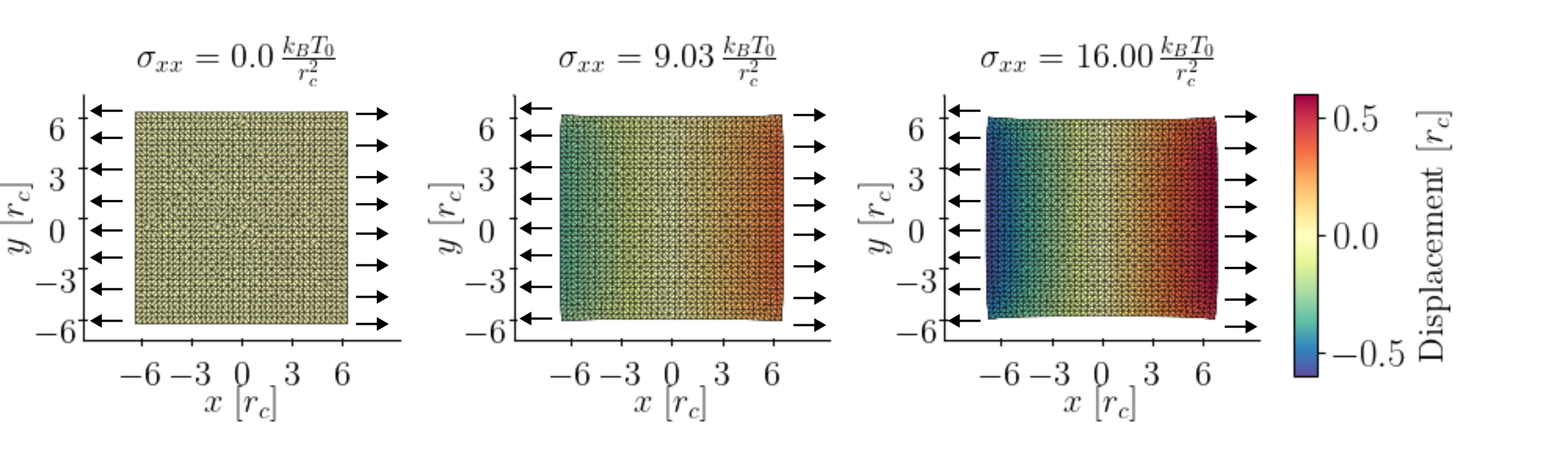}
    \caption{In-plane displacement field $u_x(x, y)$ of the rectangular sheet, stretched along the $x$ axis, for different values of dilatational stress $\sigma_{xx}$. The arrows denote the directions of the forces on the leftmost and rightmost vertices of the sheet.}
    \label{fig:field}
\end{figure}

\begin{figure}[!htbp]
    \centering
    \includegraphics[width=0.95\linewidth]{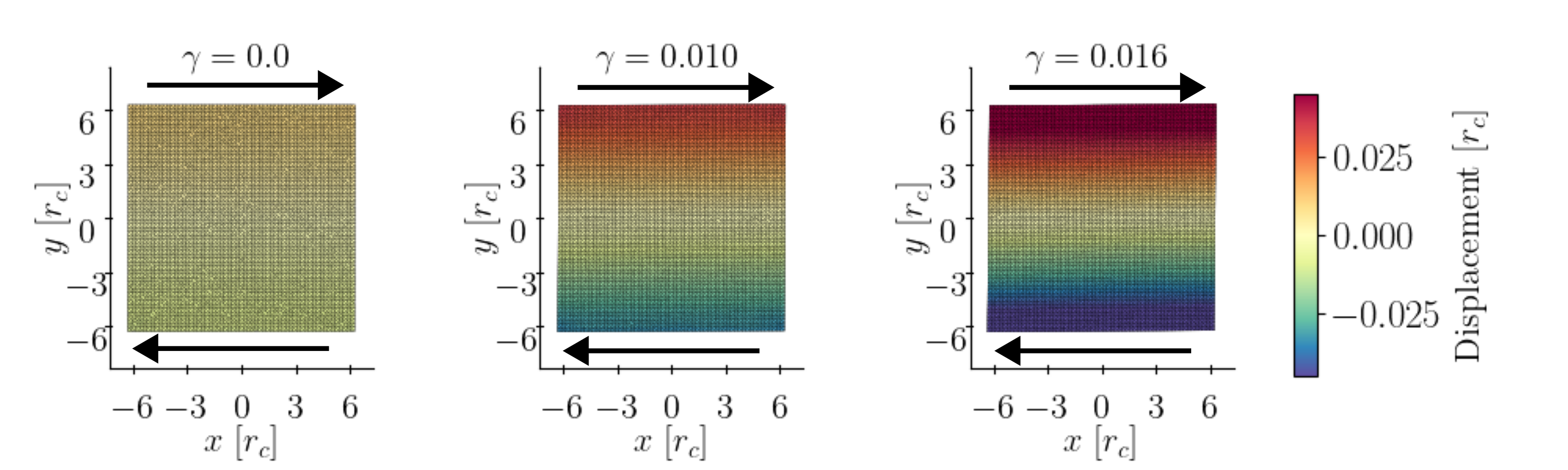}
    \caption{In-plane displacement field $u_x(x, y)$ of the rectangular sheet, for different values of shear strain $\gamma$. The arrows indicate the directions of displacement of the topmost and bottommost vertices of the sheet. }
    \label{fig:field-shear}
\end{figure}

The corresponding in-plane displacement fields $u_x(x,y)$ of the rectangular sheet are shown for different values of dilatational stress, Fig.~\ref{fig:field}, and shear strain, Fig.~\ref{fig:field-shear}.

\subsection{Microbubbles} 
\label{subsec:MBs-comp-buck}

\paragraph{Stretching}
Stretching of a microbubble results in an increased diameter along the stretching direction and a decrease of the diameter perpendicular to the stretching direction. The stretching is experimentally induced by attaching two micron-sized spherical silica beads at the ends of a given object and displacing them in opposite directions using lasers. The equilibrium condition reads \cite{evans1980}:
\begin{equation}
    \frac{\sigma_{\theta\theta}}{R_\theta} + \frac{\sigma_{\varphi \varphi}}{R_\varphi} = 0,
\end{equation}
where $R_\theta$ and $R_\varphi$ are the principal radii of curvature along the meridian and equator, respectively.
In the low strain limit $R_\theta \approx R_\varphi$, which yields $\sigma_{\varphi \varphi} = -\sigma_{\theta\theta}$. The tangential stress at the equator can be calculated from the total force: $\sigma_{\theta \theta} = \frac{F}{2\pi R}$, which yields the strain along the equator $\varepsilon_{\varphi \varphi} = \frac{F}{4\pi R \mu}$. Taking into account the relation $\varepsilon_{\varphi \varphi} = \frac{D-D_0}{D_0}$, the diameter then reads:
\begin{equation}
\label{emb_diam}
    D = D_0 - \frac{F}{2\pi \mu}.
\end{equation}

\paragraph{Compression}

Compression is achieved by exerting a uniform excess pressure $\Delta p$ on the fluid containing \textsc{emb}s. 
The induced elastic stresses within the spherical \textsc{emb} shell are homogeneous and can be determined by virtually halving the sphere and calculating the total force exerted by the pressure difference on the hemisphere, see Fig.~\ref{fig: emb-press} in the main text. This force equals $\Delta p \pi R^2$, which induces a uniform stress $\sigma_{\theta\theta} = \sigma_{\varphi\varphi}\equiv\sigma$ within the shell:
\begin{equation}
    \sigma = -\frac{\Delta p \pi R^2}{2\pi R} = -\frac{R}{2}\Delta p. 
    \label{sigmathth}
\end{equation}
The relative \textsc{emb} volume change is $\frac{\Delta V}{V} = 3\frac{\Delta R}{R} = 3\varepsilon_{\theta\theta} = 3\varepsilon_{\varphi \varphi}$, or also $\frac{\Delta V}{V}=\frac{3}{2}\frac{\Delta S}{S}$, where $\frac{\Delta S}{S}$ is the relative \textsc{emb} surface area change.
Taking into account
\begin{align}
    \varepsilon_{\theta \theta} &= \frac{\sigma_{\theta \theta}}{E} - \frac{\nu \sigma_{\varphi \varphi}}{E}  \\
    \varepsilon_{\varphi \varphi} &= -\frac{\nu\sigma_{\theta \theta}}{E} + \frac{\sigma_{\varphi \varphi}}{E},
\end{align}
or directly the definition of 2D compression
\begin{equation}
    \frac{\Delta S}{S} = \frac{1}{K_a}\frac{{\rm tr}({\bm\sigma})}{2},
\end{equation}
with ${\rm tr}({\bm\sigma}) = \sigma_{\theta \theta} + \sigma_{\varphi \varphi}$, one arrives at
\begin{equation} \label{iso-comp-sigma}
    \frac{\Delta V}{V} = \frac{3(1-\nu)}{E}\sigma = \frac{3}{2K_a}\sigma
\end{equation}
and with Eq.\,(\ref{sigmathth}) finally at the relative \textsc{emb} volume change
\begin{equation} \label{iso-comp}
    \frac{\Delta V}{V} = - \frac{3(1-\nu)R}{2E}\Delta p = -\frac{3R}{4K_a}\Delta p.
\end{equation}

For an orthotropic elastic sphere, the result Eq.\,(\ref{iso-comp-sigma}) becomes
\begin{equation}
    \frac{\Delta V}{V} = \frac{1}{E_l}\left(1 - 3\nu_{lt} + 2\frac{E_l}{E_t}\right)\sigma,
\end{equation}
which reduces to Eq.\,(\ref{iso-comp-sigma}) when $E_l = E_t = E$.

\begin{figure}[!htbp]
\centering
    \includegraphics[width=\textwidth]{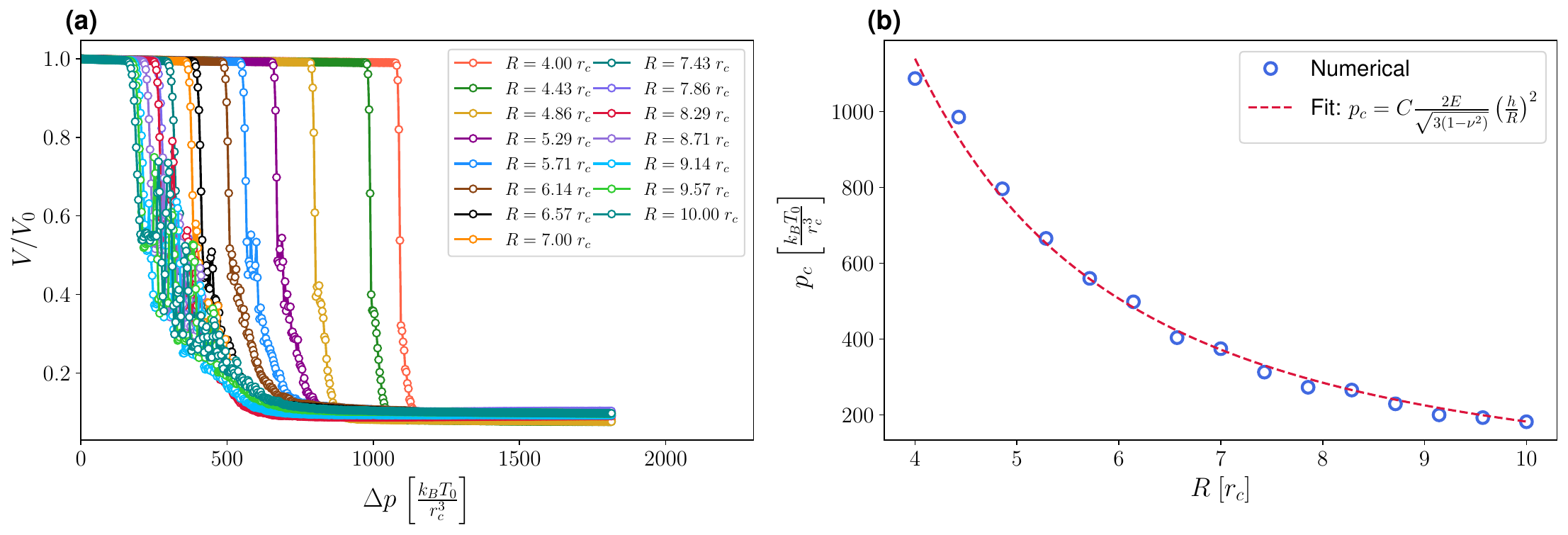}
\caption{(a) Volume of an \textsc{emb} as a function of external pressure $\Delta p$ for various \textsc{emb} sizes $R$. (b) Buckling pressure dependence on $R$, fitted (dashed red) with Eq.\,(\ref{eq: pcrit}).}
\label{fig: pcrit}
\end{figure}

\paragraph{Buckling}
Critical buckling pressure of an \textsc{emb} scales inversely with the square of its radius \cite{Zoelly:1915}:
\begin{equation}
\label{eq: pcrit}
    \Delta p_c = C\frac{2E^{3D}}{\sqrt{3(1-\nu^2)}}\left(\frac{h}{R}\right)^2,
\end{equation}
where $C$ is a dimensionless empirical factor, which in real experiments is often found to be less than 1 owing to the various imperfections of the shell. We find that the Eq.~(\ref{eq: pcrit}) is in excellent agreement with the simulations, Fig.~\ref{fig: pcrit}, with $C = 0.52 \pm 0.01$.

\subsection{Gas vesicles}
\label{subsec:GVs-comp-buck}

\paragraph{Stretching}
When stretching the \textsc{gv} along the axis one can estimate the local stress tensor in the cylindrical part from the total force: $\sigma_{zz}=\frac{F_{tot}}{2\pi R}$. The only other component is the circumferential stress $\sigma_{\varphi  \varphi}$, which is zero, $\sigma_{\varphi  \varphi} = 0$. For the resulting longitudinal and circumferential strain we therefore find rather simple relations: 
\begin{align} 
   \varepsilon_{zz} &= \frac{1}{E_l}\frac{F_{tot}}{2\pi R} \label{gv-stretch1} \\
    \varepsilon_{\varphi \varphi} &= -\frac{\nu_{lt}}{E_l}\frac{F_{tot}}{2\pi R}. \label{gv-stretch2} 
\end{align}
As has been done in the main text, one can then use these relations to estimate the longitudinal Young's modulus and the Poisson's ratio $\nu_{lt}$.

\begin{figure}[!htbp]
\centering
    \includegraphics[width=\textwidth]{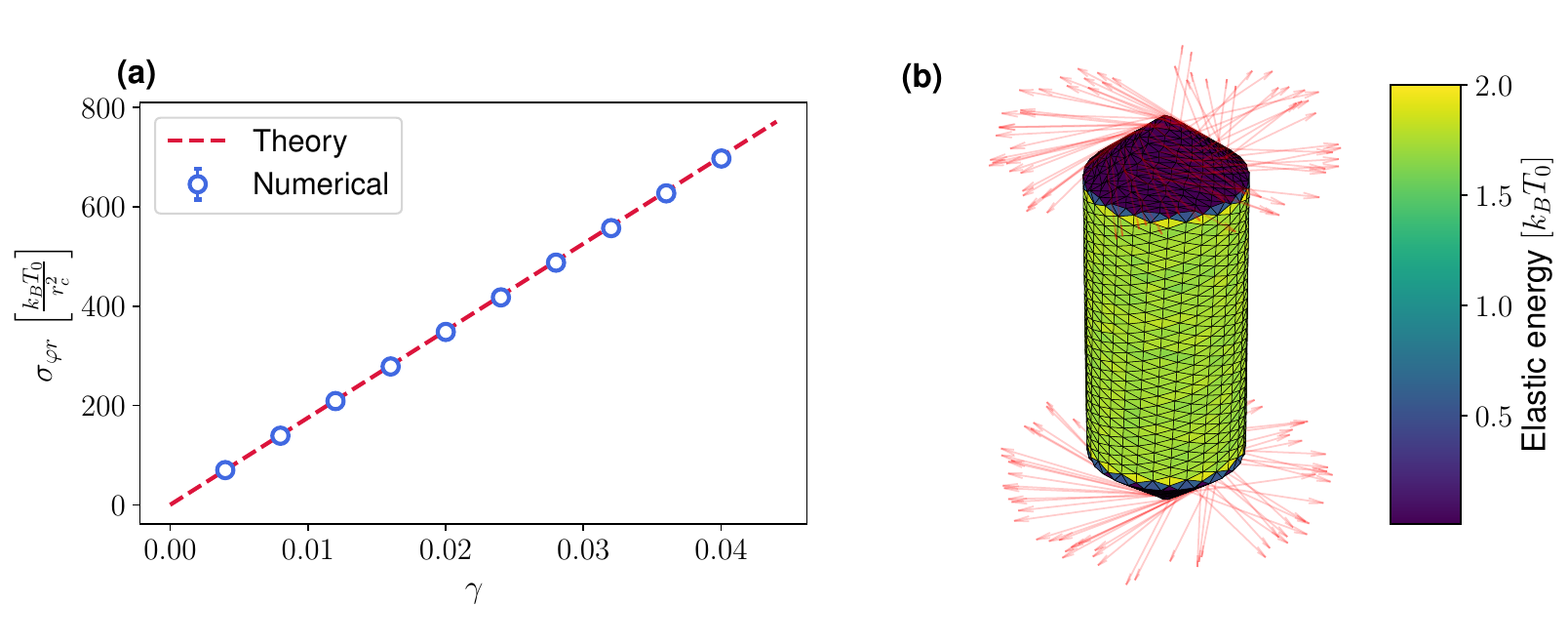}
\caption{(a) Torsional stress $\sigma_{\varphi r}$ as a function of the torsional strain $\gamma = \frac{R \theta}{H_{cyl}^0}$ for a \textsc{gv} twisted around the main axis. The numerical values are compared to theoretical values in the limit of low strain. (b) Side view of a \textsc{gv} under a specific torsional strain. The vertices are colored according to their total elastic energy.}
\label{fig: torsion_stress}
\end{figure}

\paragraph{Torsion}
On the other hand, to extract the shear modulus $G$, one has to induce shear strain on the membrane. We apply this strain by rotating the ends of the gas vesicles in opposite directions by an equal angle of $\theta/2$. The torsional strain is therefore $\gamma = \frac{R_0 \theta}{H_{cyl}^0}$. The torsional stress is calculated from the measured forces on the vertices, which are needed to sustain this strain.
As one can see from Fig.~\ref{fig: torsion_stress}, the theoretical dependence $\sigma_{\varphi r} = G \gamma$ matches the numerical calculations excellently.

\paragraph{Compression}
Similarly one can obtain an equation for the volume change of gas vesicles:
\begin{align}
    \varepsilon_{zz} &= \frac{\sigma_{zz}}{E_l} - \frac{\nu_{tl} \sigma_{\varphi \varphi}}{E_t} = \frac{\sigma_{zz}}{E_l} - \frac{\nu_{lt} \sigma_{\varphi \varphi}}{E_l} \label{epsilonzz} \\
    \varepsilon_{\varphi \varphi} &= \frac{\sigma_{\varphi \varphi}}{E_t} - \frac{\nu_{lt}\sigma_{zz}}{E_l},
\end{align}
where $\sigma_{zz} = \frac12 \Delta pR_0$, $\sigma_{\varphi \varphi} = \Delta pR_0$, and we used Eq.\,(\ref{poissonlt}). 
For cylindrical shapes, the volume change is $\frac{\Delta V}{V_0} = \varepsilon_{zz}+2\varepsilon_{\varphi \varphi}$, which yields:
\begin{equation}
    \frac{\Delta V}{V_0} = - \frac{\Delta pR_0}{2E_l}\left(1 - 4\nu_{lt} + 4\frac{E_l}{E_t}\right). \label{dVV0}
\end{equation}

\section{Dissipative Particle Dynamics (DPD)}
\subsection{Weight functions}
\label{sec:weights}

\noindent
In DPD, the force between the particles linearly decreases with interparticle distance $r_{ij}=|\mathbf{r}_{ij}|$ up to a cutoff $r_{ij}=r_c$. This is encompassed in the conservative weight function $\omega_C$ in Eq.\,(\ref{cons}):
\begin{equation}
\omega_C(r_{ij}) =  \left\{
        \begin{array}{ll}
            1-\frac{r_{ij}}{r_c}, & \quad r_{ij} < r_c \\
            0, & \quad r_{ij} \geq r_c
        \end{array}
    \right.,
    \label{omegaC}
\end{equation}

To satisfy the fluctuation-dissipation theorem, the weight functions of the random and dissipative forces Eqs.\,(\ref{diss})-(\ref{random}) are related through $\omega_R^2 = \omega_D$. Similarly, the amplitude of the random force $\sigma_{\alpha \beta}$ is fixed by temperature $T_0$ and the friction parameter $\gamma_{\alpha\beta}$ to $\sigma_{\alpha\beta}^2=2k_BT_0\gamma_{\alpha\beta}$.

At a given temperature, the viscosity of the \textsc{dpd} fluid is determined mainly by the dissipative parameter $\gamma_{\alpha\alpha}$ in Eq.\,(\ref{diss}) and the density $\rho$. To allow additional control of viscosity and the Schmidt number Eq.\,(\ref{schmidt}), several authors  \cite{fan2006, yaghoubi2015} have modified the dissipative force kernel $\omega_D$ in Eq.\,(\ref{diss}). We use the form presented in \cite{fan2006}, which is also commonly used in the red blood cell (\textsc{rbc}) modeling community \cite{Fedosov:2010b, fedosov2011ab, fedosov2011a, economides2021}:
\begin{equation} \label{eq_power}
\omega_D(r_{ij}) =  \left\{
        \begin{array}{ll}
            \left(1-\frac{r_{ij}}{r_c}\right)^{2k}, & \quad r_{ij} < r_c \\
            0, & \quad r_{ij} \geq r_c
        \end{array}
    \right., 
\end{equation}
where $k$ is the so-called kernel power. 

\subsection{Fundamental scales, simulation parameters and details}
\label{sec:compdetails}

\noindent 
The \textsc{emb} and \textsc{gv} sets of fundamental scales, described in the Methods section, are given in Table \ref{tab:scales}. 
The simulation timestep, typical simulation box sizes and \textsc{obmd} parameters used in shear flow simulations are given in Table \ref{tab-simulation}.
\begin{table}[!htb]
    \centering
    \begin{tabular}{c|c|c}\hline
        Fundamental scale & Value (\textsc{emb} unit set) & Value (\textsc{gv} unit set) \\[2ex] \hline
        Length scale $r_c$ & 0.25\,\textmu m & 35\,nm \\[2ex]
        Energy scale $\varepsilon$ & $k_BT_0$ & $k_BT_0$ \\[2ex]
        Mass scale $m$ & $5.19\times 10^{-18}$\,kg & $1.42\times 10^{-20}$\,kg \\[2ex]
        Time scale $\tau$ & $8.85\times 10^{-6}$\,s & $6.49\times 10^{-8}$\,s \\[2ex]
        Scaling factor $f_{scale}$ & $0.0074$ & $0.079$\\[2ex]
    \end{tabular}
    \caption{Fundamental scales $r_c, \varepsilon, m$, and $\tau$ for the \textsc{emb} and \textsc{gv} unit sets, along with
    the elastic coefficients down-scaling factor $f_{scale}$; $T_0 = 300\,$K.}
    \label{tab:scales}
\end{table}
\begin{table}[!htb]
    \centering
    \begin{tabular}{c|c|c}\hline
        Simulation parameter & Symbol & Value  \\[2ex] \hline
        Integration timestep & $\Delta t$ & 0.0001\,$\tau$ \\[2ex]
        Box size - mechanical simulations & $L_x\times L_y \times L_z$ & $25\,r_c \times 25\,r_c \times 25\,r_c$ \\[2ex]
        Box size - shear flow simulations & $L_x\times L_y \times L_z$ & $50\,r_c \times 25\,r_c \times 25\,r_c$ \\[2ex]
        Buffer mass control parameter & $\alpha_B$ & 0.6 \\[2ex]
        Buffer relaxation time & $\tau_B$ & $10 \Delta t$ \\[2ex]
        Buffer length & $x_B$ & $7.5\,r_c$ \\[2ex]
    \end{tabular}
    \caption{Values of the simulation parameters used, unless stated otherwise.}
    \label{tab-simulation}
\end{table}

To integrate the particle dynamics, we use a modified velocity-Verlet algorithm \cite{groot1997}:
\begin{align}
\mathbf{r}_k(t + \Delta t) &= \mathbf{r}_k(t) + \Delta t \mathbf{v}_k(t) + \frac{(\Delta t)^2}{2m_k}\mathbf{f}_k(t)\\
\mathbf{\tilde{v}}_k(t + \Delta t) &= \mathbf{v}_k(t) + \lambda \frac{\Delta t}{m_k} \mathbf{f}_k(t)\\
\mathbf{f}_k(t + \Delta t) &= \mathbf{f}_k \left( \mathbf{r}_k(t + \Delta t),  \mathbf{\tilde{v}}_k(t + \Delta t)\right) \\
\mathbf{v}_k(t + \Delta t) &= \textbf{v}_k(t) + \frac{\Delta t}{2m_k} \left( \mathbf{f}_k(t) + \mathbf{f}_k(t + \Delta t)\right),
\end{align}
with $\lambda = 1/2$, and $m_k, \mathbf{r}_k, \mathbf{v}_k$, $\mathbf{f}_k$ the mass, position vector, velocity, and total force acting on the $k$-th \textsc{dpd} bead.

\subsection{DPD interactions}
\label{sec:dpdinter}

\begin{figure}[!htbp]
    \centering
    \includegraphics[width=\textwidth]{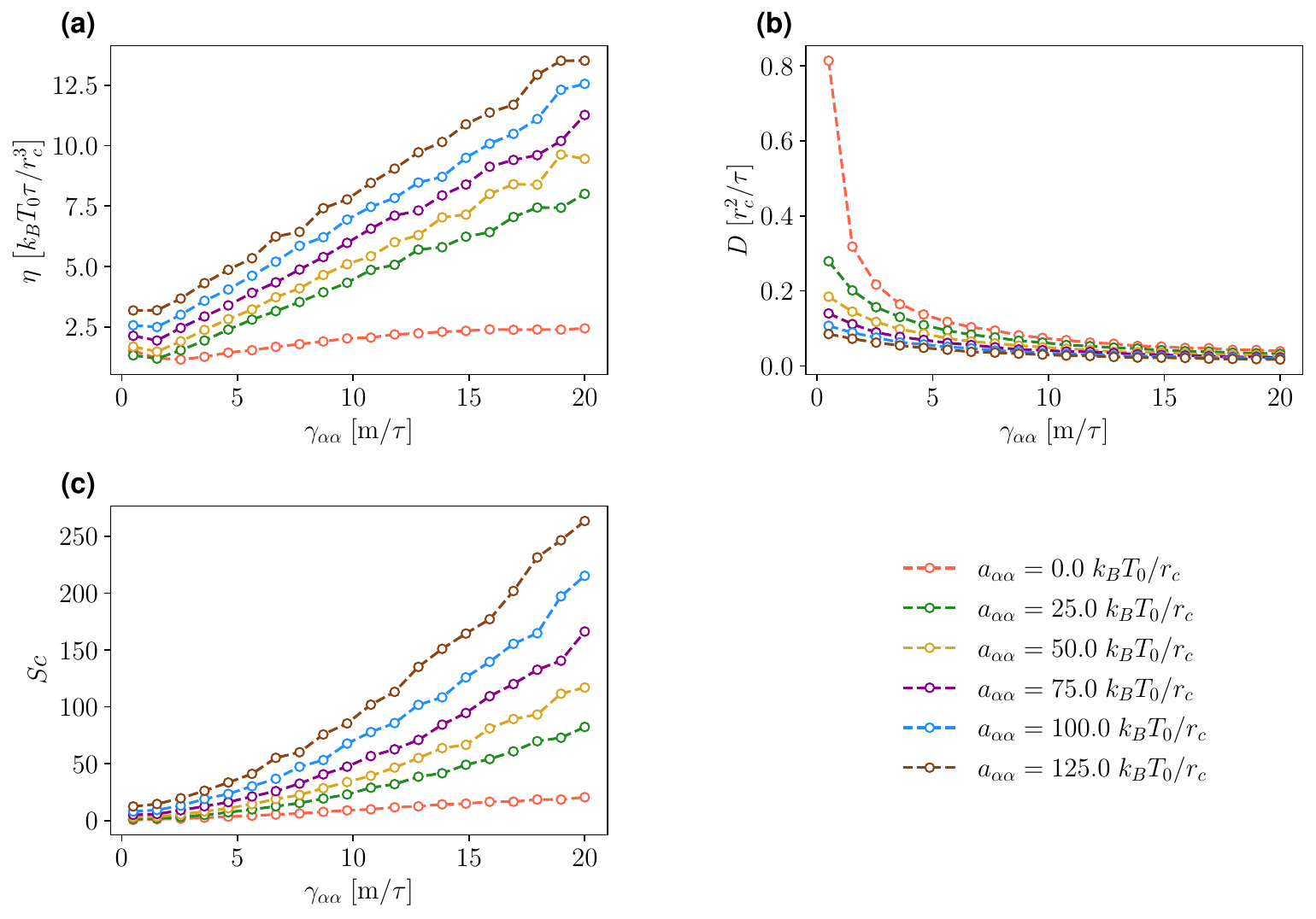}
    \caption{(a) Shear viscosity, (b) self-diffusion constant $D$ and (c) Schmidt number $Sc$ as a function of $\gamma_{\alpha\alpha}$ at  different values of the interaction parameter $a_{\alpha\alpha}$.}
    \label{fig:diffvis}
\end{figure}

\noindent
The \textsc{DPD} interactions, described by $a_{\alpha\beta}$ and $\gamma_{\alpha\beta}$ in Eqs.\,(\ref{cons})-(\ref{diss}), are given in Table \ref{tab:dpd}. These parameters are set to approximately reproduce the viscosity of the water and gas phases, as well as the no-slip boundary condition between the shell surface and the phases.
To characterize the water and gas phases, we calculate the shear viscosity $\eta$, the self-diffusion constant $D$, and the Schmidt number 
\begin{equation}
    Sc = \frac{\eta}{\rho D}
    \label{schmidt}
\end{equation}
for a relevant range of \textsc{dpd} parameters $a_{\alpha\alpha}$ and $\gamma_{\alpha\alpha}$, see Fig.~\ref{fig:diffvis}. We choose different values of parameters $\gamma_{ww}$ and $\gamma_{gg}$ depending on the length scale (Table \ref{tab:scales}) and the actual (scaled-down) viscosities of water and gas phases.

\begin{table}[!htb]
    \centering
    \begin{tabular}{c|c|c}\hline
        \textsc{dpd} parameter & Value (\textsc{emb} unit set) & Value (\textsc{gv} unit set) \\[2ex] \hline
        $a_{ww}$ & 100.0\,$\frac{k_BT_0}{r_c}$ & 100.0\,$\frac{k_BT_0}{r_c}$ \\[2ex]
        $a_{wg}$, $a_{gg}$, $a_{oo}$, $a_{og}$ & 0.0\,$\frac{k_BT_0}{r_c}$ & 0.0\,$\frac{k_BT_0}{r_c}$ \\[2ex]
        $a_{ow}$ & 40\,$\frac{k_BT_0}{r_c}$ & 40\,$\frac{k_BT_0}{r_c}$ \\[2ex] 
        $\gamma_{ww}$ & $3.5$\,$\frac{m}{\tau}$ & $18.0$\,$\frac{m}{\tau}$ \\[2ex]
        $\gamma_{gg}$ & $11.0$\,$\frac{m}{\tau}$ & $12.0$\,$\frac{m}{\tau}$ \\[2ex]
        $\gamma_{wg}$ & $0.0$\,$\frac{m}{\tau}$ & $0.0$\,$\frac{m}{\tau}$ \\[2ex]
        $\gamma_{ow}$ & $7.4$\,$\frac{m}{\tau}$ & $19.5$\,$\frac{m}{\tau}$ \\[2ex]
        $\gamma_{og}$ & $0.2$\,$\frac{m}{\tau}$ & $0.5$\,$\frac{m}{\tau}$ \\[2ex]
        $\rho_w$, $\rho_g$ & $3.0$\,$r_c^{-3}$ & $3.0$\,$r_c^{-3}$ \\[2ex]
        $k_{ww}$  & 0.25 & 0.125\\[2ex]
        $k_{gg}$  & 0.25 & 0.0\\[2ex]
        $k_{fsi}$  & 0.5 & 0.5\\[2ex]
    \end{tabular}
    \caption{Values of the \textsc{dpd} parameters used, unless stated otherwise, for \textsc{emb} and \textsc{gv} unit sets. The subscripts denote the different types of beads: ($o$)bject, ($w$)ater, ($g$)as.}\label{tab:dpd}
\end{table}

\begin{table*}[!htb]
    \centering
    \resizebox{0.8\textwidth}{!}{
    \begin{tabular}{ccc}
       \hline Parameter & Model value & Physical value \\
         \hline & \textbf{Gas vesicles} \cite{zhang2020vibration} & \\
         \hline $H=H_{cyl}+2H_{cone}$ \textsc{gv} height & $14.3$\,$r_c$ & $500$\,nm \\
          $H_{cone}$ \textsc{gv} cone height & $2.1$\,$r_c$ & $75$\,nm \\
          $D_0$ diameter & $4$\,$r_c$ & $140$\,nm \\
          $h$ shell thickness & $0.057$\,$r_c$ & $2$\,nm\\
           $\rho_s$ shell mass density & $3.97$\,$m\,r_c^{-3}$ & $1320$\,kg\,m$^{-3}$\\
           $\rho_{gas}$ gas mass density & $1.12 \times 10^{-2}$\,$m\,r_c^{-3}$ & $3.72$\,kg\,m$^{-3}$\\         
         $E^{3D}_t$ transverse Young's modulus & $1.04\times 10^7$\,$k_BT_0\,r_c^{-3}$ & $1.0$\,GPa\\        
         $E^{3D}_l$ longitudinal Young's modulus & $4.14\times 10^7$\,$k_BT_0\,r_c^{-3}$ & $4.0$\,GPa\\
         $\nu_{lt}$ Poisson's ratio & $0.3$ & $0.3$\\
        \hdashline[0.5pt/2pt]
         $K_a$ bulk modulus & $3.11 \times 10^4$\,$k_BT_0\,r_c^{-2}$ & $1.05\times 10^{-1}$\,Nm$^{-1}$\\
        $\mu$ shear modulus & $1.67 \times 10^4\,k_BT_0\,r_c^{-2}$ & $5.66\times 10^{-2}$\,Nm$^{-1}$ \\
        $\mu_L$ shear modulus & $1.67 \times 10^4\,k_BT_0\,r_c^{-2}$ & $5.66\times 10^{-2}$\,Nm$^{-1}$ \\
        $c$ anisotropic stiffness & $1.43 \times 10^5\,k_BT_0\,r_c^{-2}$ & $4.85\times 10^{-1}$\,Nm$^{-1}$ \\
       $\kappa_C$ bending parameter & $16.14\,k_BT_0$ & $6.68\times 10^{-20}\,J$\\ 
         \hline & \textbf{Protein-based microbubbles} \cite{marmottant2011buckling,ye1996sound} & \\
         \hline $R_0$ radius & $4$\,$r_c$ & $1$\,\textmu m \\
        $h$ shell thickness & $0.06$\,$r_c$ & $15$\,nm\\
         $\rho_s$ shell mass density & $3.31\,mr_c^{-3}$ & $1100$\,kg\,m$^{-3}$\\
         $\rho_{gas}$ gas mass density & $3.89\times 10^{-3}$\,$m\,r_c^{-3}$ & $1.293$\,kg\,m$^{-3}$\\                  
         $E_{3D}$ Young's modulus & $9.96\times 10^8$\,$k_BT_0\,r_c^{-3}$ & $264$\,MPa\\
         $\nu$ Poisson's ratio & $0.5$ & $0.5$\\
         \hdashline[0.5pt/2pt]
        $K_a$ bulk modulus & $4.42 \times 10^5$\,$k_BT_0\,r_c^{-2}$ & $2.93\times 10^{-2}$\,Nm$^{-1}$\\
        $\mu$ shear modulus & $1.47 \times 10^5$\,$k_BT_0\,r_c^{-2}$ & $9.77\times 10^{-3}$\,Nm$^{-1}$\\
        $\kappa_C$ bending parameter & $204.3\,k_BT_0$ & $8.46\times 10^{-19}\,J$\\              
    \end{tabular}}
    \caption{Physical (above dashed line) and model 2D (below dashed line) parameters for \textsc{gv}s and \textsc{emb}s. The 2D elastic moduli are downscaled by the scaling factor $f_{scale}$.}
    \label{tab:my_label}
\end{table*}

\subsection{Inducing pressure}
\label{sec:pressure}

\begin{figure}
    \centering
    \includegraphics[width=0.5\linewidth]{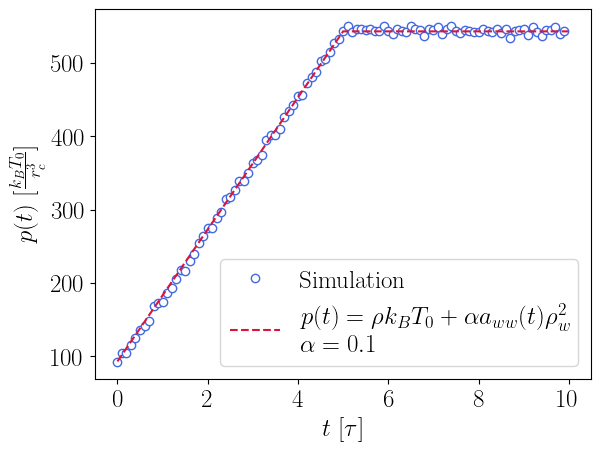}
    \caption{Pressure as a function of time (blue curve), with the theoretical prediction from the semi-empirical equation of state denoted by the red dashed line.}
    \label{fig:enter-label}
\end{figure}

\noindent
We induce pressure by increasing the interaction parameter between the water beads $a_{ww}$. 
Here, we verify that such a protocol leads to a pressure, which matches nicely to the semi-empirical equation of state Eq.\,(\ref{peos}) as seen in Fig.~\ref{fig:enter-label}.

\clearpage


\end{document}